\documentclass[prb,twocolumn,superscriptaddress,longbibliography]{revtex4-1}
\RequirePackage[mathlines]{lineno}
\usepackage{amssymb}
\usepackage{amsmath}
\usepackage{mathrsfs}
\usepackage{graphicx}
\usepackage[usenames,dvipsnames]{xcolor}
\definecolor{goodgreen}{rgb}{0.1,0.5,0}
\definecolor{goodred}{rgb}{0.7,0,0}
\usepackage[colorlinks,urlcolor=goodgreen,citecolor=blue,linkcolor=goodred]{hyperref}

\makeatletter
\newsavebox{\@brx}
\newcommand{\llangle}[1][]{\savebox{\@brx}{\(\m@th{#1\langle}\)}%
  \mathopen{\copy\@brx\kern-0.5\wd\@brx\usebox{\@brx}}}
\newcommand{\rrangle}[1][]{\savebox{\@brx}{\(\m@th{#1\rangle}\)}%
  \mathclose{\copy\@brx\kern-0.5\wd\@brx\usebox{\@brx}}}
\makeatother

\usepackage[normalem]{ulem}

\newcommand{\dcom}[1]{\textcolor{NavyBlue}{\textsf{\textbf{#1}}}}

\setpagewiselinenumbers
\modulolinenumbers[5]
\begin{document}

\preprint{\dcom{Preprint not for distribution CONFIDENTIAL, Version of \today}}
\title{Electronic band crossing in sliding bilayer graphene: Tight-binding calculations and symmetry group representation analysis}

\author{V. Nam Do}
\email{nam.dovan@phenikaa-uni.edu.vn}    
\affiliation{Department of Basic Science, Phenikaa Institute for Advanced Study (PIAS), A1 Building, Phenikaa University, Hanoi 10000, Vietnam}

\begin{abstract}
Dirac points are found to emerge due to the crossing of bands in the electronic structure of bilayer graphene for configurations in which the alignment between two hexagonal lattices preserves the parallelism of the armchair/zigzag lines between two layers. On the base of electronic calculations using a tight-binding model for the $\pi$ bands it is shown that the crossing of the energy-band dispersion curves occurs in the vicinity of the corner points of the hexagonal Brillouin zone. Group representation theory analysis confirms the emergence of such Dirac points. It is demonstrated that the band crossings at generic $\mathbf{k}$ points are guaranteed by the compatibility relations between the symmetries of eigenstates at the high-symmetry $\mathbf{k}$ points in the Brillouin zone. The presence of Dirac points governs the geometrical properties of the energy surfaces, and thus the topological structure of the Fermi energy surface and the energy spectrum. 
\end{abstract}
\maketitle

\section{Introduction}
The shape and topological structure of the Fermi energy surface are important features that govern the behavior of a material's electronic properties. \cite{Lifshitz_1960,Son_2011,Varlet_2015,Dugdale_2016} Both depend on the geometries of the energy surfaces and the crossings between them. Recently, the topic of band crossings has been extensively revisited in the context of topological characterization of the electronic structure of semimetals.\cite{Park_2017,Shao_2019,Chan_2019,Zou_2019,Hou_2021,Hirschmann_2021} It is well established that if nonsymmorphic symmetries are present in a crystalline lattice, they will enforce the crossing of electronic bands.\cite{,Young_2012,Young_2015,Wang_2016,Zhao_2016} Dirac and/or Weyl points are formed and they are globally stable.\cite{Park_2017,Shao_2019,Chan_2019,Zou_2019,Hou_2021,Hirschmann_2021} The band crossing can also occur at generic $\mathbf{k}$ points in the Brillouin zone and is independent of the symmetry properties of the system, including the lattice symmetries and the reality of the Hamiltonian. This case is called accidental band crossings that were first discussed by Herring in 1937.\cite{Herring_1930} In this case, Dirac points are formed and protected by space-time inversion.\cite{Chiu_2014,Zhao_2016,Bernevig_2016} Studying of the band crossings therefore requires not only quantitative calculations of the electronic structure, but also qualitative symmetry analysis to validate the viability of the data-driven predictions.

Dirac points are special points in the electronic structure of a material. They are nonsmooth local extremal points of energy surfaces. Their appearance usually induce saddle points. These points essentially define topological features of energy surfaces. Dirac points may be present  ``accidentally'' by Herring's means but, they are shown to be topologically protected by spatial and time symmetries.\cite{Bernevig_2013} Graphene is a typical two-dimensional material showing all such features of Dirac points. The primitive hexagonal lattice of graphene does not possess any nonsymmorphic symmetries, but Dirac points emerge from the touching of the lowest conduction energy surface and the highest valence one at the six corner ($K$) points of the Brillouin zone.\cite{Hou_2015,Berkolaiko_2018} Thus, the Fermi energy surface simply takes the form of points. The Dirac points, though stable under perturbations that do not break either time reversal or spatial inversion symmetries, may annihilate each other if the three-fold rotation is broken.\cite{Bernevig_2013,Hou_2015,Berkolaiko_2018} 
Similar to the graphene monolayer, AB-stacked bilayer graphene also possesses the Fermi energy surface comprised of points. However, the dispersion in the vicinity of these points is characterized by the parabolic law, instead of the linear law.\cite{McCann_2013,Rhokov_2016} Unlike AB-stacked bilayer graphene, AA-stacked graphene has circles surrounding each $K$ point in its Fermi energy surface. This is the result of the upward and downward shift in the energy surfaces for the two graphene layers due to the interlayer coupling under the mirror symmetry around the lattice plane (see Fig. \ref{Fig1} below). The bilayer graphene configurations with a twist angle (TBG configurations) can also have the same degree of symmetry as that of the AA- or AB-stacked configurations, depending on the value of the twist angle.\cite{Zou-2018,Yuan_2018,Kang_2018,Koshino_2018,Song_2021} However, band crossings in the TBG configurations become really complicated due to the shrinking of the Brillouin zone and the folding of the energy surfaces.\cite{Santos-2007,Shallcross-2008,Sato_2012,Rhokov_2016} 

\begin{figure*}\centering
\includegraphics[clip=true,trim=3cm 19.5cm 3cm 2cm,width=\textwidth]{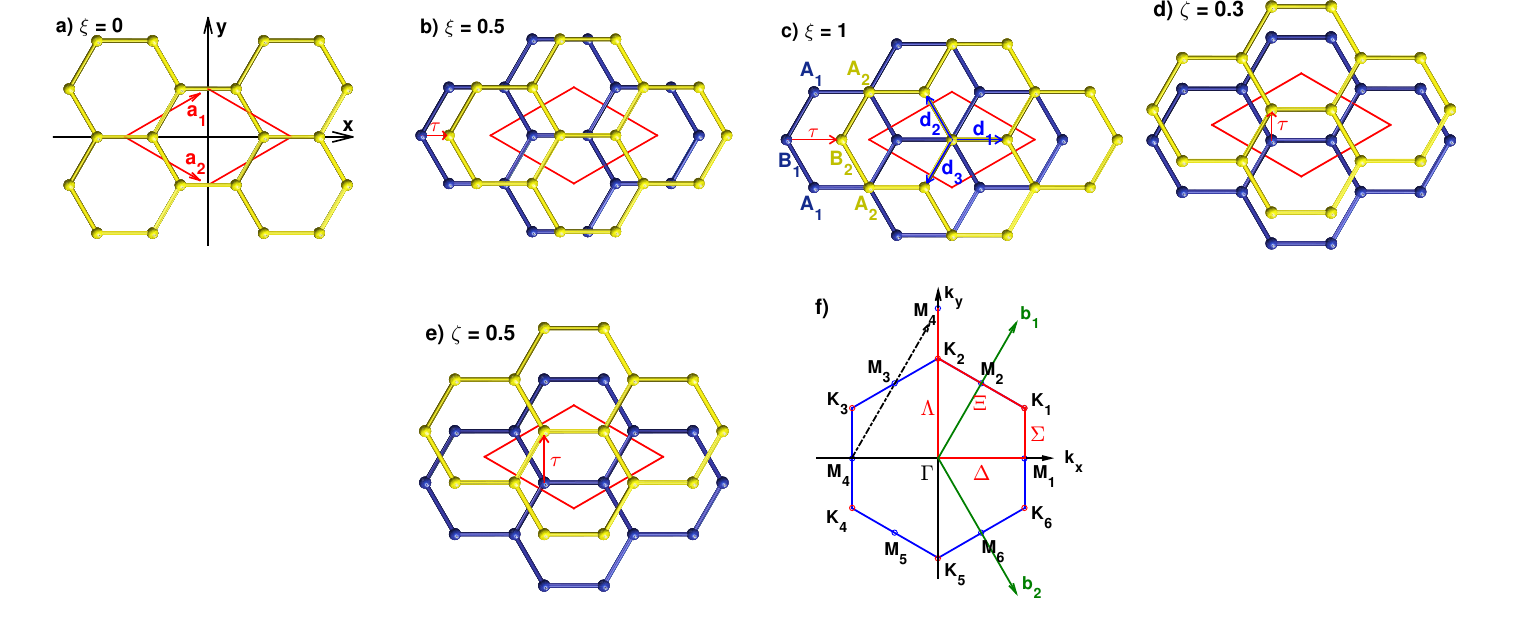}
\caption{\label{Fig1} Typical lattice configurations of the SBG system and the associated Brillouin zone. The configurations given in (a), (b) and (c) are characterized by the sliding vector $\boldsymbol{\tau} = \xi(\mathbf{a}_1+\mathbf{a}_2)/3$. The configurations given in (d) and (e) are defined by $\boldsymbol{\tau} = \zeta(\mathbf{a}_1-\mathbf{a}_2)/2$. The primitive unit cell is denoted by the red rhombus with two basis vectors $\mathbf{a}_1$ and $\mathbf{a}_2$. The three vectors $\mathbf{d}_1,\mathbf{d}_2,\mathbf{d}_2$ connect one lattice node to its three nearest in-plane neighbors. The Brillouin zone of all configurations is the same but, the symmetry of the special points $K, M$ and of the $\Lambda,\Delta,\Sigma$ axes are not because of different lattice space groups.}
\end{figure*}

Besides these special bilayer graphene configurations, there is another class in which the alignment between the two graphene lattices preserves the parallelism of the armchair/zigzag line between the two layers lines.\cite{Hibino_2009,Lin_2013,Alden_2013,Wang_2017} These structures are characterized by a sliding vector $\boldsymbol{\tau}$ and therefore are called the sliding bilayer graphene (SBG). Though having the same translation symmetry as that of the AA- and AB-stacked configurations, all rotation symmetries with the vertical axes are broken, except for some two-fold rotation axes in the lattice plane. This dramatic change in the symmetry properties leads to a significant change in the interlayer coupling, and thus the physical properties.\cite{Huang_2014,Lee_2015,Park_2015,Suszalski_2018,Huang_2019,Ho_2020,Jayaraman_2021} Electronic structure calculations for the SBG configurations suggested the presence of Dirac points, but their existence has not yet been rigorously proven.  In this paper, by using a tight-binding model for the $p_z$ electrons, we show the emergence of Dirac points in the vicinity of the corner points of the hexagonal Brillouin zone. We prove using group representation theory analysis that the emergence of such Dirac points is due to the crossing of energy dispersion curves. We will also show that the emergence of the Dirac points influences the geometrical properties of the energy surfaces, and thus the topological structure of the Fermi energy surface as well as the overall energy spectrum of the SBG configurations.

The contents of the paper are organized as follows: we first present in sub-Sec. II.A an analysis of the spatial symmetry of all bilayer configurations. We then present in sub-Sec. II.B the representation of the symmetry groups of special $\mathbf{k}$ points in a Hilbert space spanned by $4N$ electronic states of the $\pi$ bands of the system. The compatibility relations are established and presented. In sub-Sec. III.A, we present a tight-binding model for the electronic structure of the bilayer systems. In sub-Sec. III.B we present the results of the electronic structure calculations. The analysis from the symmetry group representation is used to label the energy dispersion curves. The data analysis is discussed in this subsection. Finally, the main results and conclusions are summarized in Sec. IV.

\begin{table*}[ht]
\caption{Space groups and the groups of special $\mathbf{k}$-points in the Brillouin zone of typical SBG configurations}
\label{table_I}
\centering
\begin{tabular}{llllllllllll}
\hline\hline
    Configs.   & Space group & $\Gamma$ & $K$ & $M$ & $\Delta$ & $\Lambda$ & $\Sigma$  \\ 
       \hline
    Monolayer & $P6/mmm (\#191)$ & $D_{6h}$ & $D_{3h}$ & $D_{2h}$ & $C_{2v}$ & $C_{2v}$ & $C_{2v}$ \\ 
    $\boldsymbol{\tau} = 0$ & $P6/mmm (\#191)$ & $D_{6h}$ & $D_{3h}$ & $D_{2h}$ & $C_{2v}$ & $C_{2v}$ & $C_{2v}$\\ 
    (AA-stacked)\\
    $\boldsymbol{\tau} = \frac{1}{3}(\mathbf{a}_1+\mathbf{a}_2)$  & $P\bar{3}m1 (\#164)$ & $D_{3d}$ & $D_{3}$ & $C_{2h}$ & $C_{s}$ & $C_2$ & $C_2$  \\
     (AB-stacked)\\
    $\boldsymbol{\tau} = \frac{\xi}{3}(\mathbf{a}_1+\mathbf{a}_2)^{\dagger}$ & $P2/m (\#10)$ & $C_{2h}$ & $C_2$ & $M_{1,4}: C_{2h}; M_{2,3,5,6}:C_i$ & $C_s$ & $C_2$ & $C_2$ \\
    $\boldsymbol{\tau} = \frac{1}{2}(\mathbf{a}_1\pm\mathbf{a}_2)$ & $P2/m (\#10)$ & $C_{2h}$ & $C_s$  & $M_{1,4}: C_{2h}; M_{2,3,5,6}: C_i$ & $C_2$ & $C_s$ & $C_s$ \\    
    $\boldsymbol{\tau} =  \frac{\zeta}{2}(\mathbf{a}_1-\mathbf{a}_2)^{\ddagger}$ & $P2/m (\#10)$ & $C_{2h}$ & $C_s$ & $M_{1,4}: C_{2h}; M_{2,3,5,6}: C_i$ & $C_2$ & $C_s$ & $C_s$  \\
    $\boldsymbol{\tau} \#$ & $P\Bar{1} (\#2)$ & $C_i$ & $C_1$ & $ C_i$ & $C_1$ & $C_1$ & $C_1$  \\    
    \hline\hline
    $^\dagger \xi \in (0,1)\cup (1,\frac{3}{2}); ^\ddagger \zeta \in (0,1)$
\end{tabular}
\end{table*}

\begin{table*}[t]
    \centering
    \caption{Irreducible representations at high symmetry points in the Brillouin zone of various SBG configurations}
    \label{Table_II}    
    \begin{tabular}{cccccc}
    \hline\hline
    & $\boldsymbol{\tau} = 0$ & $\boldsymbol{\tau} =\frac{\xi}{3}(\mathbf{a}_1+\mathbf{a}_2)^\dagger$ & $\boldsymbol{\tau}=\frac{1}{3}(\mathbf{a}_1+\mathbf{a}_2)$ & $\boldsymbol{\tau} =\frac{\zeta}{2}(\mathbf{a}_1-\mathbf{a}_2)^\ddagger$\\\hline
    $\Gamma$ & $A_{1g}\oplus A_{2u}\oplus B_{2g}\oplus B_{1u}$ & $2A_g\oplus 2B_u$ & $2A_{1g}\oplus2A_{2u}$ & $A_g\oplus B_u\oplus A_u\oplus B_g$ \\
    
    $K$ & $E^\prime\oplus E^{\prime\prime}$ & $2A\oplus 2B$ & $A_1\oplus A_2\oplus E$ & $2A^\prime\oplus 2A^{\prime\prime}$ \\ 
    
    $M_{1,4}$ & $B_{2g}\oplus B_{3g}\oplus A_{g}\oplus B_{1u}$ & $2A_g\oplus 2B_u$ & $2A_g\oplus 2B_u$ & $A_u\oplus B_g\oplus A_g\oplus B_u$ \\ 
    
    $M_{2,3,5,6}$ & $B_{2g}\oplus B_{3g}\oplus A_{g}\oplus B_{1u}$ & $2A_g\oplus 2A_u$ & $2A_g\oplus 2B_u$ & $2A_g\oplus 2A_u$ \\ 
    
    $\Delta $ & $A_1\oplus A_2\oplus B_1\oplus B_2$ & $4A^\prime$ & $4A^\prime$ & $2A\oplus2B$ \\     
    
    $\Lambda $ & $A_1\oplus A_2\oplus B_1\oplus B_2$ & $2A\oplus 2B$ & $2A\oplus 2B$ & $2A^\prime\oplus 2A^{\prime\prime}$ \\ 
    $\Sigma $ & $A_1\oplus A_2\oplus B_1\oplus B_2$ & $2A\oplus 2B$ & $2A\oplus 2B$ & $2A^\prime\oplus 2A^{\prime\prime}$ \\   
    \hline\hline
    \end{tabular}
\end{table*}

\begin{table*}[t]
    \centering
    \caption{Compatibility Relations between the groups $D_{6h}, D_{3d}$ and their subgroups}
    \label{Table_III}    
    \begin{tabular}{ccccc@{\hspace{1cm}}cccc@{\hspace{1cm}}ccc}
    \hline\hline\\
    $C_{2v}(\Delta,\Lambda,\Sigma)$ & $D_{6h}(\Gamma)$ & $D_{3h}(K)$ & $D_{2h}(M)$ && $C_s(\Delta)$ &$D_{3d}(\Gamma)$&$C_{2h}(M_1)$&&$C_2(\Lambda,\Sigma)$&$D_{3d}(\Gamma)$&$D_3(K_2)$\\
    \cline{1-4}\cline{6-8}\cline{10-12}\\
    $A_1$ & $A_{1g}$ & $E^\prime$ & $A_g$&&$A^\prime$&$A_{1g},A_{2u}$&$A_g,B_u$&&$A$&$A_{1g}$&$A_1,E$\\
    $A_2$ & $B_{2g}$ & $E^{\prime\prime}$ & $B_{3g}$&&&&&&$B$& $A_{2u}$&$A_2,E$\\
    $B_1$ & $B_{1u}$ & $E^{\prime}$ & $B_{2u}$\\
    $B_2$ & $A_{2u}$ & $E^{\prime\prime}$ & $B_{1u}$\\
    \hline\hline
    \end{tabular}
\end{table*}

\begin{table*}
    \centering
    \caption{Compatibility Relations of the group $C_{2h}$ and its subgroups in the SBG configurations}
    \label{Table_IV}
    \begin{tabular}{c@{\hspace{1cm}}cccc@{\hspace{1cm}}cccc@{\hspace{1cm}}ccc}
    \hline\hline\\
    $\boldsymbol{\tau}=\frac{\xi}{3}(\mathbf{a}_1+\mathbf{a}_2)$ & $C_{2}(\Lambda)$ & $C_{2h}(\Gamma)$ & $C_2(K_2)$&&$C_s(\Delta)$&$C_{2h}(\Gamma)$&$C_{2h}(M_{1})$&&$C_2(\Sigma)$&$C_{2h}(M_{1,4})$&$C_2(K_{1,2})$\\
    \cline{2-4}\cline{6-8}\cline{10-12}\\
    &$A$&$A_g$&$A$&&$A^\prime$&$A_g,B_u$&$A_g,B_u$&&$A$&$A_g$&$A$\\
    &$B$&$B_u$&$B$&&$A^{\prime\prime}$&$A_u,B_g$&$A_u,B_g$&&$B$&$B_u$&$B$\\
    \cline{2-12}\\
    $\boldsymbol{\tau}=\frac{\zeta}{2}(\mathbf{a}_1-\mathbf{a}_2)$&$C_s(\Lambda)$&$C_{2h}(\Gamma)$&$C_s(K_2)$&&$C_2(\Delta)$&$C_{2h}(\Gamma)$&$C_{2h}(M_{1,4})$&&$C_s(\Sigma)$&$C_{2h}(M_{1,4})$&$C_s(K_{1,2})$\\
    \cline{2-4}\cline{6-8}\cline{10-12}\\
    &$A^\prime$&$A_g,B_u$&$A^\prime$&&$A$&$A_g,A_u$&$A_g,A_u$&&$A^\prime$&$A_g,B_u$&$A^\prime$\\
    &$A^{\prime\prime}$&$A_u,B_g$&$A^{\prime\prime}$&&$B$&$B_g,B_u$&$B_g,B_u$&&$A^{\prime\prime}$&$A_u,B_g$&$A^{\prime\prime}$\\
    \hline\hline
\end{tabular}
\end{table*}

\begin{table*}[t]
    \centering
    \caption{Irreducible representations of symmetry groups of high symmetry points in the Brillouin zone of the $AA$- and $AB$-stacked bilayer configurations.}
    \label{Table_V}        
    \begin{tabular}{lcccl}
        \hline\hline
        Config. & $\mathbf{k}$-point & Group & Irr. Reps. & Basis vectors\\
        \hline
        $AA$-stacked & $\Gamma$ & $D_{6h}$ & $A_{1g}$ & $|A_1,p_z,0\rangle+|B_1,p_z,0\rangle-|A_2,p_z,0\rangle-|B_2,p_z,0\rangle$\\
        & & & $B_{2g}$ & $|A_1,p_z,0\rangle-|B_1,p_z,0\rangle$\\
        & & & $A_{2u}$ & $ |A_1,p_z,0\rangle+|B_1,p_z,0\rangle+|A_2,p_z,0\rangle+|B_2,p_z,0\rangle$\\
        & & & $B_{1u}$ & $|A_2,p_z,0\rangle-|B_2,p_z,0\rangle$\\    
        & $K$ & $D_{3h}$ & $E^\prime$ & $|A_1,p_z,\mathbf{K}\rangle-|A_2,p_z,\mathbf{K}\rangle;|B_1,p_z,\mathbf{K}\rangle-|B_2,p_z,\mathbf{K}\rangle$\\
        & & & $E^{\prime\prime}$ & $|A_1,p_z,\mathbf{K}\rangle+|A_2,p_z,\mathbf{K}\rangle;|B_1,p_z,\mathbf{K}\rangle+|B_2,p_z,\mathbf{K}\rangle$\\

        & $M$ & $D_{2h}$ & $A_g$ & $|A_1,p_z,\mathbf{M}\rangle+|B_1,p_z,\mathbf{M}\rangle-|A_2,p_z,\mathbf{M}\rangle-|B_2,p_z,\mathbf{M}\rangle$\\
        & & & $B_{3g}$ & $ |A_1,p_z,\mathbf{M}\rangle-|B_1,p_z,\mathbf{M}\rangle+|A_2,p_z,\mathbf{M}\rangle-|B_2,p_z,\mathbf{M}\rangle$\\
        & & & $B_{1u}$  & $|A_1,p_z,\mathbf{M}\rangle+|B_1,p_z,\mathbf{M}\rangle+|A_2,p_z,\mathbf{M}\rangle+|B_2,p_z,\mathbf{M}\rangle$\\
        & & & $B_{2u}$ & $|A_1,p_z,\mathbf{M}\rangle-|B_1,p_z,\mathbf{M}\rangle-|A_2,p_z,\mathbf{M}\rangle-|B_2,p_z,\mathbf{M}\rangle$\\
        
        & $\Lambda$ & $C_{2v}$ & $A_1$ & $ |A_1,p_z,\mathbf{k}_{\Lambda}\rangle+|B_1,p_z,\mathbf{k}_{\Lambda}\rangle-|A_2,p_z,\mathbf{k}_{\Lambda}\rangle-|B_2,p_z,\mathbf{k}_{\Lambda}\rangle$\\
        & & & $A_2$ & $ |A_1,p_z,\mathbf{k}_{\Lambda}\rangle-|B_1,p_z,\mathbf{k}_{\Lambda}\rangle+|A_2,p_z,\mathbf{k}_{\Lambda}\rangle-|B_2,p_z,\mathbf{k}_{\Lambda}\rangle$\\
        & & & $B_1$ & $ |A_1,p_z,\mathbf{k}_{\Lambda}\rangle+|B_1,p_z,\mathbf{k}_{\Lambda}\rangle-|A_2,p_z,\mathbf{k}_{\Lambda}\rangle-|B_2,p_z,\mathbf{k}_{\Lambda}\rangle$\\
        & & & $B_2$ & $ |A_1,p_z,\mathbf{k}_{\Lambda}\rangle+|B_1,p_z,\mathbf{k}_{\Lambda}\rangle+|A_2,p_z,\mathbf{k}_{\Lambda}\rangle+|B_2,p_z,\mathbf{k}_{\Lambda}\rangle$\\

        & $\Sigma$ & $C_{2v}$ & $A_1$ & $ |A_1,p_z,\mathbf{k}_{\Sigma}\rangle+e^{i\frac{2\pi}{3}}|B_1,p_z,\mathbf{k}_{\Sigma}\rangle-|A_2,p_z,\mathbf{k}_{\Sigma}\rangle-e^{-i\frac{2\pi}{3}}|B_2,p_z,\mathbf{k}_{\Sigma}\rangle$\\
        & & & $A_2$ & $ |A_1,p_z,\mathbf{k}_{\sigma}\rangle-e^{i\frac{2\pi}{3}}|B_1,p_z,\mathbf{k}_{\Sigma}\rangle-|A_2,p_z,\mathbf{k}_{\Sigma}\rangle+e^{i\frac{-2\pi}{3}}|B_2,p_z,\mathbf{k}_{\Sigma}\rangle$\\
        & & & $B_1$ & $ |A_1,p_z,\mathbf{k}_{\sigma}\rangle+e^{i\frac{2\pi}{3}}|B_1,p_z,\mathbf{k}_{\Sigma}\rangle-|A_2,p_z,\mathbf{k}_{\Sigma}\rangle-e^{-i\frac{2\pi}{3}}|B_2,p_z,\mathbf{k}_{\Sigma}\rangle$\\
        & & & $B_2$ & $ |A_1,p_z,\mathbf{k}_{\sigma}\rangle+e^{i\frac{2\pi}{3}}|B_1,p_z,\mathbf{k}_{\Sigma}\rangle+|A_2,p_z,\mathbf{k}_{\Sigma}\rangle+e^{i\frac{-2\pi}{3}}|B_2,p_z,\mathbf{k}_{\Sigma}\rangle$\\
        
        & $\Delta$ & $C_{2v}$ & $A_1$ & $ |A_1,p_z,\mathbf{k}_\Delta\rangle+|B_1,p_z,\mathbf{k}_\Delta\rangle-|A_2,p_z,\mathbf{k}_\Delta\rangle-|B_2,p_z,\mathbf{k}_\Delta\rangle$\\
        & & & $A_2$ & $|A_1,p_z,\mathbf{k}_\Delta\rangle-|B_1,p_z,\mathbf{k}_\Delta\rangle+|A_2,p_z,\mathbf{k}_\Delta\rangle-|B_2,p_z,\mathbf{k}_\Delta\rangle$\\
        & & & $B_1$  & $|A_1,p_z,\mathbf{k}_\Delta\rangle-|B_1,p_z,\mathbf{k}_\Delta\rangle-|A_2,p_z,\mathbf{k}_\Delta\rangle+|B_2,p_z,\mathbf{k}_\Delta\rangle$\\
        & & & $B_2$ & $|A_1,p_z,\mathbf{k}_\Delta\rangle+|B_1,p_z,\mathbf{k}_\Delta\rangle+|A_2,p_z,\mathbf{k}_\Delta\rangle+|B_2,p_z,\mathbf{k}_\Delta\rangle$\\        

        $AB$-stacked & $\Gamma$ & $D_{3d}$ & $A_{1g}$ & $|A_1,p_z,0\rangle -|B_2,p_z,0\rangle$\\
        & & & $A_{1g}$ & $|B_1,p_z,0\rangle - |A_2,p_z,0\rangle$\\
        & & & $A_{2u}$ & $|A_1,p_z,0\rangle + |B_2,p_z,0\rangle$\\
        & & & $A_{2u}$ & $|B_1,p_z,0\rangle + |A_2,p_z,0\rangle$\\
        
        & $K$ & $D_3$ & $A_1$ & $|B_1,p_z,\mathbf{K}\rangle-|A_2,p_z,\mathbf{K}\rangle$\\
        & & & $A_2$ & $|A_1,p_z,\mathbf{K}\rangle+|B_2,p_z,\mathbf{K}\rangle$\\
        & & & $E$ & $|A_1,p_z,\mathbf{K}\rangle; |B_2,p_z,\mathbf{K}\rangle$\\
        
        & $M$ & $C_{2h}$ & $A_g$ & $|B_1,p_z,\mathbf{M}\rangle-|A_2,p_z,\mathbf{M}\rangle$\\
        & & & $A_g$ & $|A_1,p_z,\mathbf{M}\rangle - |B_2,p_z,\mathbf{M}\rangle$\\
        & & & $B_u$ & $|B_1,p_z,\mathbf{M}\rangle+|A_2,p_z,\mathbf{M}\rangle$\\
        & & & $B_u$ & $|A_1,p_z,\mathbf{M}\rangle+|B_2,p_z,\mathbf{M}\rangle$\\
        & $\Lambda,\Sigma$ & $C_2$ & $A$ & $|A_1,p_z,\mathbf{k}_{\Lambda,\Sigma}\rangle-|B_2,p_z,\mathbf{k}_{\Lambda,\Sigma}\rangle$\\
        & & & $A$ & $|B_1,p_z,\mathbf{k}_{\Lambda,\Sigma}\rangle-|A_2,p_z,\mathbf{k}_{\Lambda,\Sigma}\rangle$\\
        & & & $B$ & $|A_1,p_z,\mathbf{k}_{\Lambda,\Sigma}\rangle+|B_2,p_z,\mathbf{k}_{\Lambda,\Sigma}\rangle$\\
        & & & $B$ & $|B_1,p_z,\mathbf{k}_{\Lambda,\Sigma}\rangle+|A_2,p_z,\mathbf{k}_{\Lambda,\Sigma}\rangle$\\
        
        & $\Delta$ & $C_s$ & $A^\prime$ & $|A_1,p_z,\mathbf{k}_{\Delta}\rangle$\\
        & & & $A^\prime$ & $|B_1,p_z,\mathbf{k}_{\Delta}\rangle$\\
        & & & $A^\prime$ & $|A_2,p_z,\mathbf{k}_{\Delta}\rangle$\\
        & & & $A^\prime$ & $|B_2,p_z,\mathbf{k}_{\Delta}\rangle$\\        
        \hline\hline
    \end{tabular}
\end{table*}

\begin{table*}[t]
    \centering
    \caption{Irreducible representations of the symmetry groups of high symmetry points in the Brillouin zone of the $SBG$ configurations with $\boldsymbol{\tau}=\xi(\mathbf{a}_1+\mathbf{a}_2)/3$.}
    \label{Table_VI}        
    \begin{tabular}{lcccl}
        \hline\hline
        Config. & $\mathbf{k}$-point & Group & Irr. Reps. & Basis vectors\\
        \hline
        $\boldsymbol{\tau}=\frac{\xi}{3}(\mathbf{a}_1+\mathbf{a}_2)$ & $\Gamma$ & $C_{2h}$ & $A_g$ & $|A_1,p_z,0\rangle-|B_2,p_z,0\rangle$\\
        & & & $A_g$ & $|B_1,p_z,0\rangle-|A_2,p_z,0\rangle$\\
        & & & $B_u$ & $|A_1,p_z,0\rangle+|B_2,p_z,0\rangle$\\
        & & & $B_u$ & $|B_1,p_z,0\rangle+|A_2,p_z,0\rangle$\\
        
        & $K$ & $C_{2}$ & $A$ & $|A_1,p_z,\mathbf{K}\rangle-|B_2,p_z,\mathbf{K}\rangle$\\
        & & & $A$ & $|B_1,p_z,\mathbf{K}\rangle-|A_2,p_z,\mathbf{K}\rangle$\\
        & & & $B$ & $|A_1,p_z,\mathbf{K}\rangle+|B_2,p_z,\mathbf{K}\rangle$\\
        & & & $B$ & $|B_1,p_z,\mathbf{K}\rangle+|A_2,p_z,\mathbf{K}\rangle$\\
        
        & $M_{1,4}$ & $C_{2h}$ & $A_g$ & $|A_1,p_z,\mathbf{M}_{1,4}\rangle-e^{-i\frac{2\pi}{3}}e^{i\xi\pi}|B_2,p_z,\mathbf{M}_{1,4}\rangle$\\
        & & & $A_g$ & $|B_1,p_z,\mathbf{M}_{1,4}\rangle-e^{i\frac{2\pi}{3}}e^{i\xi\pi}|A_2,p_z,\mathbf{M}_{1,4}\rangle$\\
        & & & $B_u$ & $|A_1,p_z,\mathbf{M}_{1,4}\rangle+e^{-i\frac{2\pi}{3}}e^{i\xi\pi}|B_2,p_z,\mathbf{M}_{1,4}\rangle$\\
        & & & $B_u$ & $|B_1,p_z,\mathbf{M}_{1,4}\rangle+e^{i\frac{2\pi}{3}}e^{i\xi\pi}|A_2,p_z,\mathbf{M}_{1,4}\rangle$\\
        
        & $M_{2,3,5,6}$ & $C_i$ & $A_g$ & $|A_1,p_z,\mathbf{M}_{2,3,5,6}\rangle-e^{-i\frac{\pi}{3}}e^{i\xi\frac{\pi}{2}}|B_2,p_z,\mathbf{M}_{2,3,5,6}\rangle$\\
        & & & $A_g$ & $|B_1,p_z,\mathbf{M}_{2,3,5,6}\rangle-e^{i\frac{\pi}{3}}e^{i\xi\frac{\pi}{2}}|A_2,p_z,\mathbf{M}_{2,3,5,6}\rangle$\\
        & & & $A_u$ & $|A_1,p_z,\mathbf{M}_{2,3,5,6}\rangle+e^{-i\frac{\pi}{3}}e^{i\xi\frac{\pi}{2}}|B_2,p_z,\mathbf{M}_{2,3,5,6}\rangle$\\
        & & & $A_u$ & $|B_1,p_z,\mathbf{M}_{2,3,5,6}\rangle+e^{i\frac{\pi}{3}}e^{i\xi\frac{\pi}{2}}|A_2,p_z,\mathbf{M}_{2,3,5,6}\rangle$\\
        
        & $\Lambda,\Sigma$ & $C_2$ & $A$ & $|A_1,p_z,\mathbf{k}_{\Lambda,\Sigma}\rangle-|B_2,p_z,\mathbf{k}_{\Lambda,\Sigma}\rangle$\\ 
        & & & $A$ & $|B_1,p_z,\mathbf{k}_{\Lambda,\Sigma}\rangle-|A_2,p_z,\mathbf{k}_{\Lambda,\Sigma}\rangle$\\ 
        & & & $B$ & $|A_1,p_z,\mathbf{k}_{\Lambda,\Sigma}\rangle+|A_2,p_z,\mathbf{k}_{\Lambda,\Sigma}\rangle$\\ 
        & & & $B$ & $|B_1,p_z,\mathbf{k}_{\Lambda,\Sigma}\rangle+|A_2,p_z,\mathbf{k}_{\Lambda,\Sigma}\rangle$\\         
        
        & $\Delta$ & $C_s$ & $A^{\prime}$ & $|A_1,p_z,\mathbf{k}_{\Delta}\rangle$\\ 
        & & & $A^{\prime}$ & $ |B_1,p_z,\mathbf{k}_{\Delta}\rangle$\\ 
        & & & $A^{\prime}$ & $|A_2,p_z,\mathbf{k}_{\Delta}\rangle$\\
        & & & $A^{\prime}$ & $|B_2,p_z,\mathbf{k}_{\Delta}\rangle$\\ 
        \hline\hline
    \end{tabular}
\end{table*}

\begin{table*}[t]
    \centering
    \caption{Irreducible representations of the symmetry groups of high symmetry points in the Brillouin zone of the $SBG$ configurations with $\boldsymbol{\tau}=\zeta(\mathbf{a}_1-\mathbf{a}_2)/2$.}
    \label{Table_VII}        
    \begin{tabular}{lcccl}
        \hline\hline
        Config. & $\mathbf{k}$-point & Group & Irr. Reps. & Basis vectors\\
        \hline
        $\boldsymbol{\tau}=\frac{\zeta}{2}(\mathbf{a}_1-\mathbf{a}_2)$ & $\Gamma$ & $C_{2h}$ & $A_g$ & $|A_1,p_z,0\rangle+|B_1,p_z,0\rangle-|A_2,p_z,0\rangle-|B_2,p_z,0\rangle$\\
        & & & $B_g$ & $|A_1,p_z,0\rangle-|B_1,p_z,0\rangle+|A_2,p_z,0\rangle-|B_2,p_z,0\rangle$\\
        & & & $A_u$ & $|A_1,p_z,0\rangle-|B_1,p_z,0\rangle-|A_2,p_z,0\rangle+|B_2,p_z,0\rangle$\\
        & & & $B_u$ & $|A_1,p_z,0\rangle+|B_1,p_z,0\rangle+|A_2,p_z,0\rangle+|B_2,p_z,0\rangle$\\

        & $K$ & $C_s$ & $A^\prime$ & $|A_1,p_z,\mathbf{K}\rangle+|B_1,p_z,\mathbf{K}\rangle$\\
        & & & $A^\prime$ & $|A_2,p_z,\mathbf{K}\rangle+|B_2,p_z,\mathbf{K}\rangle$\\
        & & & $A^{\prime\prime}$ & $|A_1,p_z,\mathbf{K}\rangle-|B_1,p_z,\mathbf{K}\rangle$\\
        & & & $A^{\prime\prime}$ & $|A_2,p_z,\mathbf{K}\rangle-|B_1,p_z,\mathbf{K}\rangle$\\
        
        & $M_{1,4}$ & $C_{2h}$ & $A_g$ & $|A_1,p_z,\mathbf{M}_{1,4}\rangle+e^{i\frac{2\pi}{3}}|B_1,p_z,\mathbf{M}_{1,4}\rangle-|A_2,p_z,\mathbf{M}_{1,4}\rangle-e^{i\frac{2\pi}{3}}|B_2,p_z,\mathbf{M}_{1,4}\rangle$\\
        & & & $B_g$ & $|A_1,p_z,\mathbf{M}_{1,4}\rangle-e^{i\frac{2\pi}{3}}|B_1,p_z,\mathbf{M}_{1,4}\rangle+|A_2,p_z,\mathbf{M}_{1,4}\rangle-e^{i\frac{2\pi}{3}}|B_2,p_z,\mathbf{M}_{1,4}\rangle$\\
        & & & $A_u$ & $|A_1,p_z,\mathbf{M}_{1,4}\rangle-e^{i\frac{2\pi}{3}}|B_1,p_z,\mathbf{M}_{1,4}\rangle-|A_2,p_z,\mathbf{M}_{1,4}\rangle+e^{i\frac{2\pi}{3}}|B_2,p_z,\mathbf{M}_{1,4}\rangle$\\
        & & & $B_u$ & $|A_1,p_z,\mathbf{M}_{1,4}\rangle+e^{i\frac{2\pi}{3}}|B_1,p_z,\mathbf{M}_{1,4}\rangle+|A_2,p_z,\mathbf{M}_{1,4}\rangle+e^{i\frac{2\pi}{3}}|B_2,p_z,\mathbf{M}_{1,4}\rangle$\\
        
        & $M_{2,3,5,6}$ & $C_i$ & $A_g$ & $|A_1,p_z,\mathbf{M}_{2,3,5,6}\rangle-e^{i\frac{\pi}{3}}e^{-i\zeta\frac{\pi}{2}}|B_2,p_z,\mathbf{M}_{2,3,5,6}\rangle$\\
        & & & $A_g$ & $|B_1,p_z,\mathbf{M}_{2,3,5,6}\rangle-e^{-i\frac{\pi}{3}}e^{i\zeta\frac{\pi}{2}}|A_2,p_z,\mathbf{M}_{2,3,5,6}\rangle$\\
        & & & $B_u$ & $|A_1,p_z,\mathbf{M}_{2,3,5,6}\rangle+e^{i\frac{\pi}{3}}e^{-i\zeta\frac{\pi}{2}}|B_2,p_z,\mathbf{M}_{2,3,5,6}\rangle$\\
        & & & $B_u$ & $|B_1,p_z,\mathbf{M}_{2,3,5,6}\rangle+e^{-i\frac{\pi}{3}}e^{i\zeta\frac{\pi}{2}}|A_2,p_z,\mathbf{M}_{2,3,5,6}\rangle$\\   
        
        & $\Lambda,\Sigma$ & $C_s$ & $A^\prime$ & $|A_1,p_z,\mathbf{k}_{\Lambda,\Sigma}\rangle+|B_1,p_z,\mathbf{k}_{\Lambda,\Sigma}\rangle$\\ 
        & & & $A^\prime$ & $|A_2,p_z,\mathbf{k}_{\Lambda,\Sigma}\rangle+|B_2,p_z,\mathbf{k}_{\Lambda,\Sigma}\rangle$\\ 
        & & & $A^{\prime\prime}$ & $|A_1,p_z,\mathbf{k}_{\Lambda,\Sigma}\rangle-|B_1,p_z,\mathbf{k}_{\Lambda,\Sigma}\rangle$\\ 
        & & & $A^{\prime\prime}$ & $|A_2,p_z,\mathbf{k}_{\Lambda,\Sigma}\rangle-|B_2,p_z,\mathbf{k}_{\Lambda,\Sigma}\rangle$\\         
        
        & $\Delta$ & $C_2$ & $A$ & $|A_1,p_z,\mathbf{k}_{\Delta}\rangle-|A_2,p_z,\mathbf{k}_{\Delta}\rangle$\\   
        & & & $A$ & $|B_1,p_z,\mathbf{k}_{\Delta}\rangle-|B_2,p_z,\mathbf{k}_{\Delta}\rangle$\\   
        & & & $B$ & $|A_1,p_z,\mathbf{k}_{\Delta}\rangle+|A_2,p_z,\mathbf{k}_{\Delta}\rangle$\\   
        & & & $B$ & $|B_1,p_z,\mathbf{k}_{\Delta}\rangle+|B_2,p_z,\mathbf{k}_{\Delta}\rangle$\\                   
        \hline\hline
    \end{tabular}
\end{table*}

\section{Lattice configurations}
\subsection{Real space symmetry}
Let us consider a system of two flat graphene layers stacked together with the interlayer distance $d_{GG} = 3.35$ \AA. We do not consider the relative twisting between the two layers, but only the sliding among them, which is characterized by a sliding vector $\boldsymbol{\tau}$. Accordingly, when $\boldsymbol{\tau}=0$ the SBG configuration corresponds to the AA-stacked configuration. Because the sliding does not break the parallel property in the lattice plane, the resulted complex atomic lattices always have the same translation symmetry of the AA-stacked configuration with the primitive cell defined by two basis vectors:
\begin{equation}
\mathbf{a}_1 = \frac{a}{2}(\sqrt{3}\hat{\mathbf{x}}+\hat{\mathbf{y}});
\mathbf{a}_2 = \frac{a}{2}(\sqrt{3}\hat{\mathbf{x}}-\hat{\mathbf{y}}),\label{Eq1}
\end{equation}
where $a=\sqrt{3}a_{CC}$ is the lattice constant and $a_{CC} = 1.45$ \AA\, is the distance between two nearest lattice sites; $\hat{\mathbf{x}}$ and $\hat{\mathbf{y}}$ denote unit vectors in the Cartesian coordinate frame. The reciprocal lattice is then built by the following two basis vectors: 
\begin{equation}
\mathbf{b}_1 = \frac{2\pi}{\sqrt{3}a}(\hat{\mathbf{x}}+\sqrt{3}\hat{\mathbf{y}}); 
\mathbf{b}_2 = \frac{2\pi}{\sqrt{3}a}(\hat{\mathbf{x}}-\sqrt{3}\hat{\mathbf{y}}).\label{Eq2}
\end{equation}
These vectors define a Brillouin zone that is shaped as a hexagon with six corner points, called the $K$ points. They are determined by:
\begin{subequations}
\begin{align}
\mathbf{K}_1 &=\frac{2}{3}\mathbf{b}_1+\frac{1}{3}\mathbf{b}_2 = -\mathbf{K}_4,\label{Eq3a}\\
\mathbf{K}_2 &=\frac{1}{3}\mathbf{b}_1-\frac{1}{3}\mathbf{b}_2 = -\mathbf{K}_5,\label{Eq3b}\\
\mathbf{K}_6 &=\frac{1}{3}\mathbf{b}_1+\frac{2}{3}\mathbf{b}_2 = -\mathbf{K}_3.\label{Eq3c}
\end{align}
\end{subequations}

The SBG lattices always have the spatial inversion centers. One of these is the central point of the parallelogram $A_1A_2B_2B_1$, i.e., at the point determined by the vector $(\mathbf{A_1A_2}+\mathbf{A_1B_1})/2 = (\boldsymbol{\tau}+\mathbf{d}_1)/2$; here $\mathbf{d}_1 = (\mathbf{a}_1+\mathbf{a}_2)/3$. We choose the origin $O$ at this spatial inversion center, except for the case of the AA-stacked configuration (i.e., SBG with $\boldsymbol{\tau} = 0$) in which $O$ is chosen to be at the highest symmetry central point of one of the hexagonal atomic rings. Accordingly, the positions of the carbon atoms is determined by the vector $\mathbf{R}_\alpha = \mathbf{R}+\mathbf{d}_\alpha, \alpha = A_1, B_1, A_2,B_2$, where:
\begin{subequations}
\begin{align}
\mathbf{d}_{A_1} &= \mathbf{OA}_1 =  -\frac{1}{2}(\mathbf{d}_1+\boldsymbol{\tau}),\label{Eq4a}\\
\mathbf{d}_{B_1} &= \mathbf{OB}_1 = +\frac{1}{2}(\mathbf{d}_1-\boldsymbol{\tau}),\label{Eq4b}\\
\mathbf{d}_{A_2} &= \mathbf{OA}_2 = -\frac{1}{2}(\mathbf{d}_1-\boldsymbol{\tau}),\label{Eq4c}\\
\mathbf{d}_{B_2} &= \mathbf{OB}_2 = +\frac{1}{2}(\mathbf{d}_1+\boldsymbol{\tau}),\label{Eq4d}
\end{align}
\end{subequations}

Due to the periodicity of the real space lattices, it is necessary to consider the sliding vector $\boldsymbol{\tau}$ in the domain of a triangle. Accordingly, we consider sliding in two directions:  $\boldsymbol{\tau}$ parallel to $\mathbf{d}_1$ and $\boldsymbol{\tau}$ perpendicular to $\mathbf{d}_1$. Figure 1 shows the schema for some typical SBG configurations along with the previously described vectors and other useful notation. Among all possible configurations of the SBG systems, the AA-stacked configuration with $\boldsymbol{\tau} = 0$ has the highest symmetry and can be described using the symmorphic space group $P6/mmm (\#191)$. Accordingly, the lattice has a mirror plane $M_{xy}$ lying in the middle between two graphene layers; a 6-fold rotation axis $C_{6z}$ perpendicular to $M_{xy}$; 3 mirror planes contain the $C_{6z}$ rotation axis and crossing the middle points of the hexagonal edges; 3 other mirror planes also contain the axis $C_{6z}$ and two corner points of the hexagonal ring. The quotient group of $P6/mmm$ is exactly the point group $D_{6h}$ with 24 symmetry operations, which are classified into 12 equivalent classes. The symmetry operations of the $D_{6h}$ point group are specified in our notation as $D_{6h} =  \left\{E,2C_{6z}\right.$, $2C_{3z}$, $C_{2z}$, $3C_{2x}$, $3C_{2y}$, $i$, $2S_{3z}$, $2S_{6z}$, $M_{xy}$, $3M_{yz}$, $\left. 3M_{xz}\right\}$.

The AB-stacked configuration has lower symmetry and is obtained when the sliding vector $\boldsymbol{\tau} = \mathbf{d}_1$. In this configuration, the plane $xy$ is no longer a mirror symmetry plane. The principle rotation axis is the 3-fold $C_{3z}$, going through one atomic site in the top graphene layer and another in the second layer. Additionally, three out of six planes containing the $C_{3z}$ axis and going through the middle points of the hexagonal ring are no longer mirror planes. The other three planes going through the corner points of the atomic ring remain being mirror planes. The space group of the lattice is symmorphic and is denoted by $P\bar{3}m1 (\# 164)$. Its quotient group is exactly the point group $D_{3d}$ with 12 symmetry operations. There are 6 equivalent classes in the group and the symmetry operations in our notation are $D_{3d} = \{E,2C_{3z}, 3C_{2y}, i, 2S_{6z}, 3M_{xz}\}$.

When sliding along the direction of the vector $\mathbf{d}_1$, i.e., $\boldsymbol{\tau} = \xi(\mathbf{a}_1+\mathbf{a}_2)/3$, where $\xi \in (0,1)\cup (1, 3/2)$, the resulting configurations have much lower symmetry. They are in the space group of $P2/m (\# 10)$. This is a symmorphic group whose the quotient group is identical to the point group $C_{2h}$. The symmetry operations of $C_{2h}$ include $C_{2h} = \{E,C_{2y}, i, M_{xz}\}$. When $\xi = 1$ we obtain the AB-stacked configuration. When $\xi = 3/2$ then $\boldsymbol{\tau} = (\mathbf{a}_1 + \mathbf{a}_2)/2$ and we obtain a special SBG configuration, see Fig. 1(e). The space group of this  configuration is found to be $P222 (\#16)$, which is also a symmorphic group. It has the associated point group $D_2$ which consists of 4 symmetry operations, i.e., $D_2 = \{E, C_{2z}, C_{2y}, C_{2x}\}$. However, it should noted that the $D_2$ group is isomorphic to the groups $C_{2h}$ and  $C_{2v} = \{E, C_{2z}, M_{yz}, M_{xz}\}$. We chose to work with the $C_{2h}$ group.

When the sliding vector is along the direction perpendicular to the $\mathbf{d}_1$ vector, $\boldsymbol{\tau}\propto (\mathbf{a}_1-\mathbf{a}_2)/2$, the resulting SBG configurations have the space group $P2/m (\#10)$. The associated point group is $C_{2h} = \{E, C_{2x}, i, M_{yz}\}$. We find that the SBG configuration with $\boldsymbol{\tau} = (\mathbf{a}_1-\mathbf{a}_2)/2$ is identical to the one with $\boldsymbol{\tau} = (\mathbf{a}_1+\mathbf{a}_2)/2$. 

For the other values of the sliding vector $\boldsymbol{\tau}$ the resulting atomic lattices have the lowest symmetry. They fall under the space group $P\Bar{1}$, which is the product of the translation group and the point group $C_i = \{E, i\}$ consisting of only two symmetry operations, the identity and the spatial inversion. In summary, for all possible sliding vectors the space groups of the resulting SBG configurations are symmorphic, i.e., always decomposed into the product of a point group and the translation group with the Braivais lattice vectors. The symmetry groups of special $\mathbf{k}$ points in the first Brillouin zone of several typical SBG configurations are presented in Table 1. One should note that despite the first Brillouin zone of all SBG configurations adopting the same hexagonal shape, the symmetry groups of the $K$- and $M$-points are different from one configuration to the other because of their different space groups.

\subsection{Electronic structure symmetries}
Electronic states in a crystalline atomic lattice are expressed in terms of Bloch functions. In the rhombus unit cell of all SBG configurations, there are 4 distinct carbon atoms, named $A_1,B_1, A_2$ and $B_2$. In this work, we only consider energy bands formed by the hybridization of the $p_z$ orbitals of each carbon atom. The ket vector $|\alpha,p_z,\mathbf{R}+\mathbf{d}_\alpha\rangle$ stands for the orbital $p_z$ of the carbon atom $\alpha$ located at the position $\mathbf{d}_\alpha$ in the unit cell $\mathbf{R}$, i.e., $\phi_{\alpha,p_z}(\mathbf{r}-\mathbf{R}-\mathbf{d}_\alpha)\rightarrow |\alpha,p_z,\mathbf{R+d}_\alpha\rangle$. The set of Bloch vectors $\{|\alpha,p_z,\mathbf{k}\rangle\,|\, \forall \alpha = A_1,B_1,A_2,B_2;\mathbf{k}\in \text{Brillouin zone (BZ)}\}$, wherein
\begin{equation}
    |\alpha,p_z,\mathbf{k}\rangle = \frac{1}{\sqrt{N}}\sum_\mathbf{R}e^{-i\mathbf{k}\cdot(\mathbf{R}+\mathbf{d}_\alpha)}|\alpha,p_z,\mathbf{R+d}_\alpha\rangle,\label{Eq5}
\end{equation}
therefore forming a set of basis vectors that represents all the  electronic states. We use this basis set to realize a representation of the space group and of the $\mathbf{k}$-vector groups of the atomic lattices. Formally, $T_g$ denotes a linear operator representing a symmetry operation $g$ from the lattice symmetry group. Its action on the basis vector is formally defined by:\cite{Malard_2009,Kogan_2012,Kogan_2014,Soares_2014}
\begin{equation}
    T_g|\alpha,p_z,\mathbf{k}\rangle = |g\alpha,gp_z,g\mathbf{k}\rangle,\label{Eq6}
\end{equation}
where $g\alpha\rightarrow \alpha^\prime, gp_z\rightarrow \pm p_z$ and $g\mathbf{k}\rightarrow \mathbf{k}^\prime$. From Eq. (\ref{Eq5}) one should note that $|\alpha,p_z,\mathbf{k}+\mathbf{G}\rangle = e^{i\mathbf{G}\cdot\mathbf{d}_\alpha}|\alpha,p_z,\mathbf{k}\rangle$ where $\mathbf{G}$ is any reciprocal lattice vector. We use Eq. (\ref{Eq6}) to determine all character values of the representation. The results for all points and axes in the first Brillouin zone are presented in Table I and Table II.

In the case of the AA-stacked configuration, it has six $K$ points with $D_{3h}$ symmetry, six $M$ points with $D_{2h}$ symmetry, and three high symmetry axes ($\Delta,\Lambda$ and $\Sigma$) with $C_{2v}$ symmetry. Representing these symmetry groups in the 4-dimensional Hilbert space we obtain the following results.  For the $\Gamma$ point, the representation of the $D_{6h}$ group is reducible and therefore decomposed into four 1-dimension irreducible representations $A_{1g}\oplus A_{2u}\oplus B_{2g}\oplus B_{1u}$. For the $K$ points, the representation of the $D_{3h}$ group is decomposable into two 2-dimensional irreducible representations, $E^\prime\oplus E^{\prime\prime}$. For the $M$ points, the representation of the $D_{2h}$ group is decomposable into four 1-dimensional representations $A_{g}\oplus B_{1g}\oplus B_{2u}\oplus B_{3g}$. 

For the AB-stacked configuration, which has lower symmetry compared to the AA-stacked configuration, all six $K$ and all $M$ points of the Brillouin zone have the same symmetry groups $D_3$ and $C_{2h}$, respectively. However, unlike for the AA configuration, the three high symmetry axes  $\Lambda,\Sigma$ and $\Delta$ do not have the same symmetry group. The $\Lambda$ and $\Sigma$ axes have the same $C_2$ group whereas the $\Delta$ axis has $C_s$ symmetry. Representing these symmetry groups in the 4-dimensional Hilbert space it can be seen that at the $\Gamma$ point, the representation of the $D_{3d}$ group is decomposable into two 1-dimensional irreducible representations $A_{1g}$ and two other 1-dimensional ones $A_{2u}$, i.e., $2A_{1g}\oplus 2A_{2u}$. For the $K$ points, the representation of the $D_{3}$ group can be decomposed into two 1-dimensional and one 2-dimensional irreducible representations, i.e., $A_1\oplus A_2\oplus E$. For the $M$ points, the representation of the $C_{2h}$ group decomposes into two 1-dimensional irreducible representations $A_g$ and two other 1-dimensional irreducible representations $B_u$, i.e., $2A_g\oplus 2B_u$.

For the SBG configurations with $\boldsymbol{\tau} = \xi(\mathbf{a}_1+\mathbf{a}_2)/3$ such that $\xi\neq 0$ or $2/3$, they belong to the $P2/m$ space group. The symmetry group of the $\Gamma$ point is $C_{2h}$. We verified that all six $K$ points belong to the same $C_2$ symmetry group. However, this is not the case for the $M$ points. The two points, $M_1$ and $M_4$, belong to the same $C_{2h}$ symmetry group whereas $M_{2,3,5,6}$ points are of the lower $C_i$ symmetry group (a group with only two symmetry operations: the identity and the spatial inversion). We also examined that the two axes, $\Lambda$ and $\Sigma$, have the same $C_2$ symmetry, unlike the $\Delta$ axis which has $C_s$ symmetry. For the configurations under discussion, the representation of the $C_{2h}$ group for the $\Gamma$ and $M_{1,4}$ points is a reducible representation. It is decomposable into two 1-dimensional irreducible representations $A_g$ (even under the inversion) and two other 1-dimensional irreducible representations $B_u$ (odd under the inversion):  $2A_g\oplus 2B_u$. For the  $K$ points, the representation of the group $C_2$ is decomposable into two 1-dimensional representations $A$ and two other 1-dimensional representations: $B$, i.e., $2A\oplus 2B$. For the points $M_{2,3,5,6}$ the representation of the group $C_i$ may be decomposed into $2A_g\oplus 2A_u$. 

For the SBG configurations with $\boldsymbol{\tau} = \zeta(\mathbf{a}_1-\mathbf{a}_2)/2$, wherein $\xi\neq 0$ and $1$, although they have the same $P2/m$ space group as the previously mentioned SBG configurations (i.e.,  $\boldsymbol{\tau} = \xi(\mathbf{a}_1+\mathbf{a}_2)/3$), the six $K$ points are of $C_s$ symmetry. Furthermore, the $M_1$ and $M_4$ points are of $C_{2h}$ symmetry, while the $M_{2,3,5,6}$ points belong to $C_i$ symmetry. For the $\Lambda$ and $\Sigma$ axes, they are described by the $C_s$ symmetry group, and the $\Delta$ axis by the $C_2$ symmetry group. The representation of the group of the $\Gamma$ and $M_{1,4}$ points is decomposable into four one-dimensional irreducible representations $A_g\oplus B_g\oplus A_u\oplus B_u$. The representation of the $C_s$ group for the $K$ points is decomposable into the following 1-dimensional irreducible representations $2A^\prime\oplus 2A^{\prime\prime}$.

In addition to analyzing the representation of the symmetry groups of the high symmetry $K$ and $M$ points in the Brillouin zone we also established the compatibility relations of such  points along the high symmetry axes. These results are presented in Tables III and IV for all typical configurations of the bilayer system.

The analysis of the representation of symmetry groups of the $\mathbf{k}$ points in the Brillouin zone does not provide quantitative information on the energy eigenvalues of the Bloch-Hamiltonian matrix. However, it allows one to determine the basis vectors for the Hilbert subspaces of  the irreducible representations of the symmetry groups. We determined the vectors spanning such invariant subspaces for all the relevant symmetry groups. We present these results in Tables V, VI and VIII. The unnormalized basis vectors of the invariant subspaces are expressed as the linear combinations of the Bloch vectors of the $p_z$ electrons in the $A_1,B_1,A_2,B_2$ sublattices of the total bilayer lattices. Remarkably, for generic SBG configurations, we find that the combination coefficients in the basis vectors spanning the subspaces representing the symmetry groups of the $M$ points depend on the sliding vector $\boldsymbol{\tau}$, while this is not the case for the $\Gamma$ and $K$ points. 

In Fig. \ref{Fig2}, we present the energy dispersion curves along our previous identified symmetry axes that connect the  high symmetry points in the Brillouin zone. The four dispersion curves are labeled by the names of the irreducible representations of the symmetry groups of the corresponding $\mathbf{k}$ axes. In next section, we will discuss this qualitative analysis along with numerical results obtained from a tight-binding model for the electrons localized in the $p_z$ atomic orbitals of the bilayer graphene system.

\section{Electronic structure calculations}
\subsection{Tight-binding model}
In the basis set of the atomic orbital vectors  $\{|\alpha,p_z,\mathbf{R}+\mathbf{d}_\alpha\rangle\,|\,\forall \alpha=A_1,B_1,A_2,B_2;\mathbf{R}\in \text{Bravais lattice}\}$, the tight-binding Hamiltonian is defined as:
\begin{eqnarray}\label{Eq7}
H&=&\sum_{\mathbf{R},\mathbf{R}_j}\sum_{\alpha,\beta}t(\mathbf{d}_{\alpha\beta}^j)|\alpha,p_z,\mathbf{R}+\mathbf{d}_\alpha\rangle\langle\beta,p_z,\mathbf{R}_j+\mathbf{d}_\beta|,
\end{eqnarray}
where $\mathbf{d}_{\alpha\beta}^j = \mathbf{R}_j+\mathbf{d}_\beta-(\mathbf{R}+\mathbf{d}_\alpha)$ are the vectors connecting the site $\alpha$ to the neighbor sites $\beta$; $t(\mathbf{d}_{\alpha\beta}^j)$ refers to the electron hopping integral between the $p_z$ orbitals of the neighboring atomic sites. The value of this quantity depends only on the distance between the sites, $d^j_{\alpha\beta} = |\mathbf{d}^j_{\alpha\beta}|$. For numerical calculations we use the following model for the hopping integral:
\begin{equation}\label{Eq8}
    t(\mathbf{d}_{\alpha\beta}^j) = V_{pp\pi}(d^j_{\alpha\beta})\sin^2\theta^z_{\alpha\beta}+V_{pp\sigma}(d^j_{\alpha\beta})\cos^2\theta^z_{\alpha\beta},
\end{equation}
where $\cos\theta^z_{\alpha\beta} = (\mathbf{d}^j_{\alpha\beta}\cdot\mathbf{e}_z)/d_{\alpha\beta}^j$ and
\begin{subequations}
\begin{eqnarray}
V_{pp\pi}(d^j_{\alpha\beta}) &=& V^0_{pp\pi}\exp\left(-\frac{d^j_{\alpha\beta}-a_{CC}}{r_0}\right),\label{Eq9a}\\
V_{pp\sigma}(d^j_{\alpha\beta}) &=& V^0_{pp\sigma}\exp\left(-\frac{d^j_{\alpha\beta}-a_{CC}}{r_0}\right).\label{Eq9b}
\end{eqnarray}
\end{subequations}
In this model, the parameters are commonly set to $V^0_{pp\pi} = -2.7$ eV, $V_{pp\sigma}^0 = 0.48$ eV, and $r_0 = 0.148a$.\cite{Moon-2013,Koshino-2015,Le-2018,Le-2019,Do-2019}

Now, expanding the vector $|\alpha,p_z,\mathbf{R}+\mathbf{d}_\alpha\rangle$ in terms of the Fourier transform of the Bloch vector $|\alpha,p_z,\mathbf{k}\rangle$ yields:
\begin{equation}\label{Eq10}
    |\alpha,p_z,\mathbf{R}+\mathbf{d}_\alpha\rangle=\frac{1}{\sqrt{N}}\sum_{\mathbf{k}}e^{i\mathbf{k}\cdot(\mathbf{R}+\mathbf{d}_\alpha)}|\alpha,p_z,\mathbf{k}\rangle.
\end{equation}
Substitute this into Eq. (\ref{Eq7}) we obtain the following expression for the Hamiltonian:
\begin{equation}\label{Eq11}
    H = \sum_{\mathbf{k}}\sum_{\alpha,\beta}|\alpha,p_z,\mathbf{k}\rangle h_{\alpha\beta}(\mathbf{k})\langle\beta,p_z,\mathbf{k}|,
\end{equation}
where 
\begin{eqnarray}\label{Eq12}
h_{\alpha\beta}(\mathbf{k}) &=& \sum_{j}t(\mathbf{d}^j_{\alpha\beta})e^{-i\mathbf{k}\cdot\mathbf{d}_{\alpha\beta}^j}.
\end{eqnarray}
Eq. (\ref{Eq11}) is the result of the basis transformation of the Hamiltonian operator from the localized $p_z$ atomic orbitals basis set to the basis set of Bloch vectors $\{|\alpha,p_z,\mathbf{k}\rangle |\forall \alpha = A_1,B_1,A_2,B_2; \mathbf{k}\in BZ\}$. By defining a field vector $|\Psi_{p_z,\mathbf{k}}\rangle = (|A_1,p_z,\mathbf{k}\rangle,|B_1,p_z,\mathbf{k}\rangle,|A_2,p_z,\mathbf{k}\rangle,|B_2,p_z,\mathbf{k}\rangle)^T$ we can arrange the coefficients $h_{\alpha\beta}(\mathbf{k})$ into a matrix, called the Bloch-Hamiltonian matrix $h(\mathbf{k})$, which is:
\begin{equation}\label{Eq13}
    h(\mathbf{k}) = \left(\begin{array}{cccc}
    0&f_\mathbf{k}& u_\mathbf{k}&v_\mathbf{k}\\
    f^*_\mathbf{k}&0&w_\mathbf{k}&u_\mathbf{k}\\
    u^*_\mathbf{k}&w^*_\mathbf{k}&0&f_\mathbf{k}\\
    v^*_\mathbf{k}&u^*_\mathbf{k}&f^*_\mathbf{k}&0
    \end{array}\right)
\end{equation}
with elements:
\begin{subequations}
\begin{eqnarray}
f_\mathbf{k} &=& \sum_{j}t(\mathbf{d}^j_{A_1B_1})e^{-i\mathbf{k}\cdot\mathbf{d}^j_{A_1B_1}}\label{Eq14a},\\
u_\mathbf{k} &=& \sum_{j}t(\mathbf{d}^j_{A_1A_2})e^{-i\mathbf{k}\cdot\mathbf{d}^j_{A_1A_2}}\label{Eq14b},\\
v_\mathbf{k} &=& \sum_{j}t(\mathbf{d}^j_{A_1B_2})e^{-i\mathbf{k}\cdot\mathbf{d}^j_{A_1B_2}}\label{Eq14c},\\
w_\mathbf{k} &=& \sum_{j}t(\mathbf{d}^j_{B_1A_2})e^{-i\mathbf{k}\cdot\mathbf{d}^j_{B_1A_2}}\label{Eq14d}.
\end{eqnarray}
\end{subequations}
The index $j$ in the above equations runs over all the atomic sites neighboring a central site. In our calculation we approximate the in-plane electronic coupling to be only between $p_z$ orbitals of nearest neighbors. Hence, Eq.(\ref{Eq14a}) can be expressed as
\begin{equation}\label{Eq15}
    f_\mathbf{k} = t_0\left[e^{-ik_xa_{CC}}+2e^{i\frac{k_xa_{CC}}{2}}\cos\left(\frac{\sqrt{3}k_ya_{CC}}{2}\right)\right] 
\end{equation}
where $t_0=t_{A_1B_1}=t_{A_2B_2}=t(\mathbf{d}_1)$. For the interlayer electronic coupling, we consider the coupling of one $p_z$ orbital in one layer only to the $p_z$ orbitals in the other layer that are within the vicinity of the cutoff radius $R^{c} = \sqrt{d_{GG}^2+a^2_{CC}}$. One may notice that the Bloch-Hamiltonian matrix,  $h(\mathbf{k})$, as defined by Eq. (\ref{Eq13}) is not periodic with the reciprocal lattice vector $\mathbf{G}= m\mathbf{b}_1+n\mathbf{b}_2$. However, by changing the  Bloch vectors basis set, $\{|\alpha,p_z,\mathbf{k}\rangle\}$, to the new ones $\{|\alpha,p_z,\mathbf{d}_\alpha,\mathbf{k}\rangle\}$ where $|\alpha,p_z,\mathbf{d}_\alpha,\mathbf{k}\rangle = e^{i\mathbf{k}\cdot\mathbf{d}_\alpha}|\alpha,p_z,\mathbf{k}\rangle$, one can obtain a new form of the Bloch-Hamitlonian matrix $H(\mathbf{k})$. This form satisfies the periodicity condition $H(\mathbf{k}+\mathbf{G})=H(\mathbf{k})$. Both $h(\mathbf{k})$ and $H(\mathbf{K})$ result in the identical energy spectrum for a system since they are related to each other by a unitary transformation.

\begin{figure*}\centering
\includegraphics[clip=true,trim=1.5cm 6.5cm 2cm 7cm,width=0.9\columnwidth]{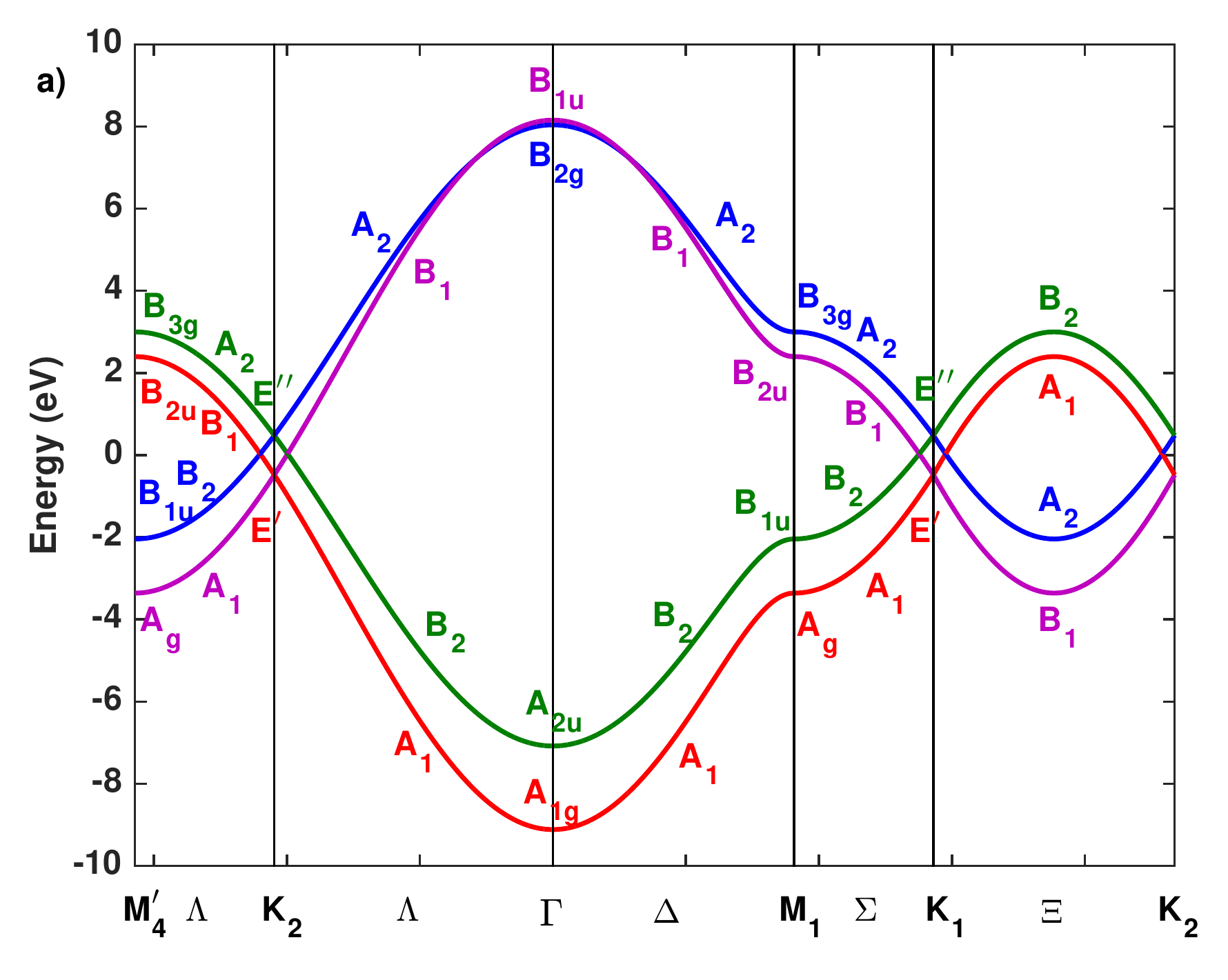}
\includegraphics[clip=true,trim=1.5cm 6.5cm 2cm 7cm,width=0.9\columnwidth]{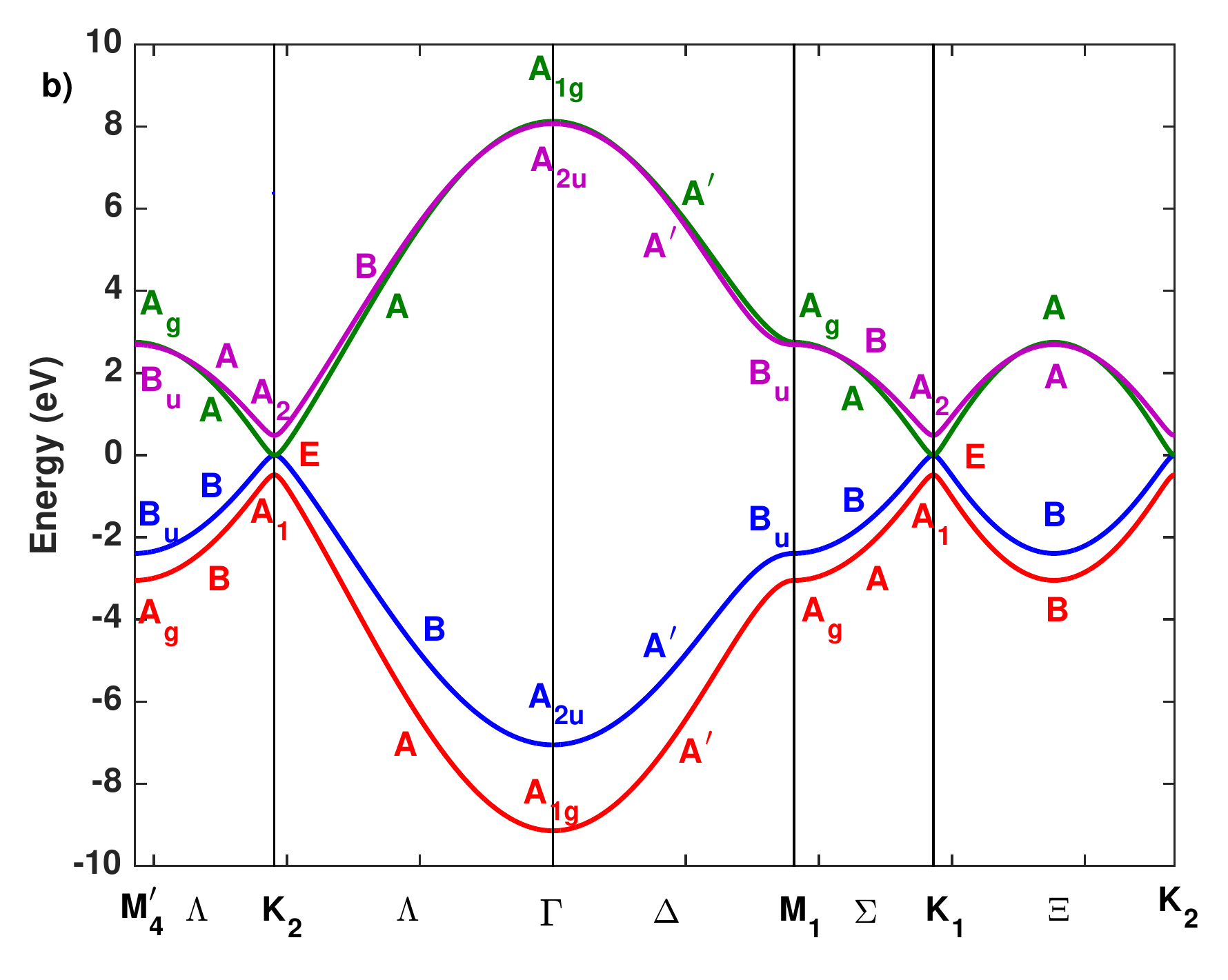}
\includegraphics[clip=true,trim=1.5cm 6.5cm 2cm 7cm,width=0.9\columnwidth]{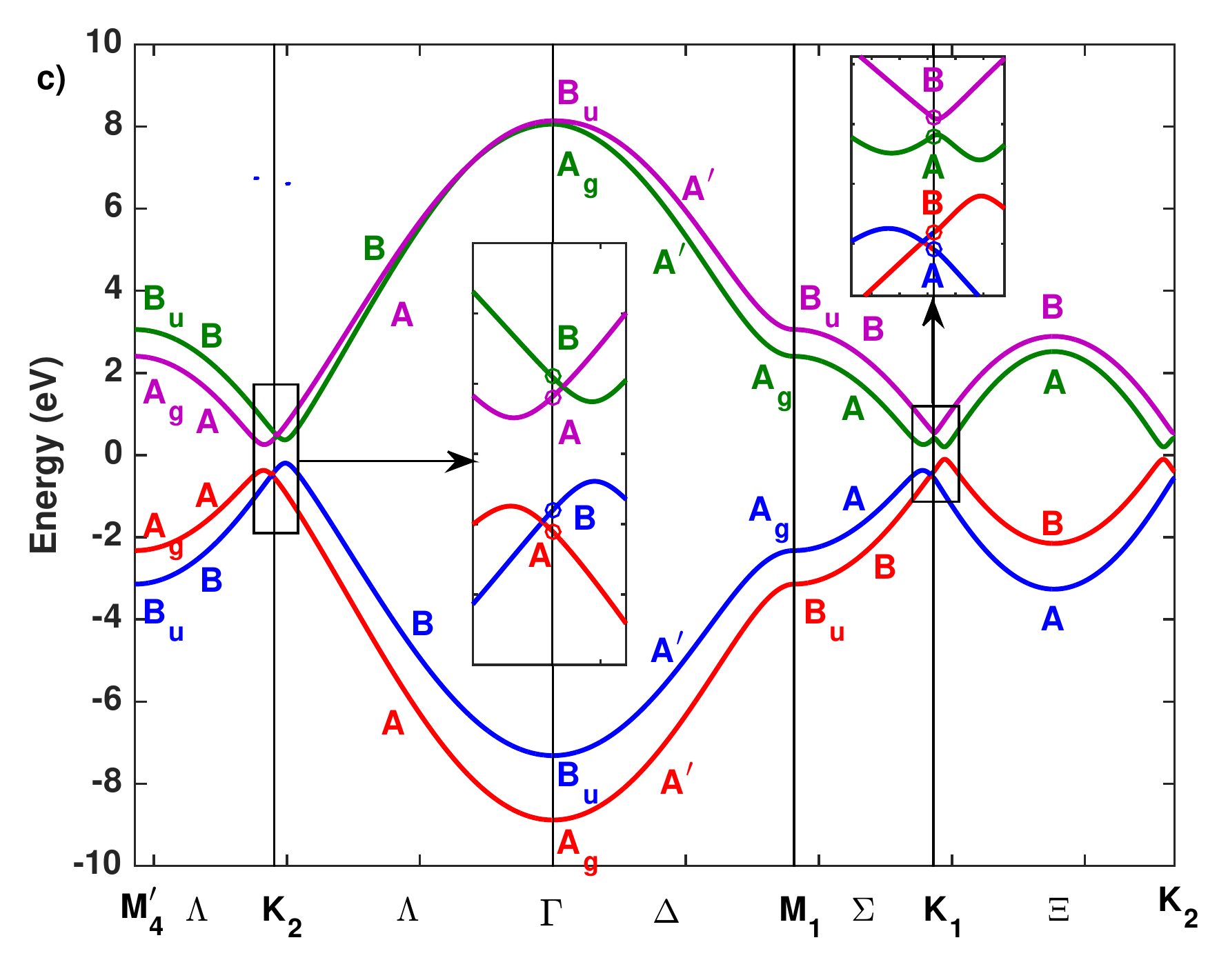}
\includegraphics[clip=true,trim=1.5cm 6.5cm 2cm 7cm,width=0.9\columnwidth]{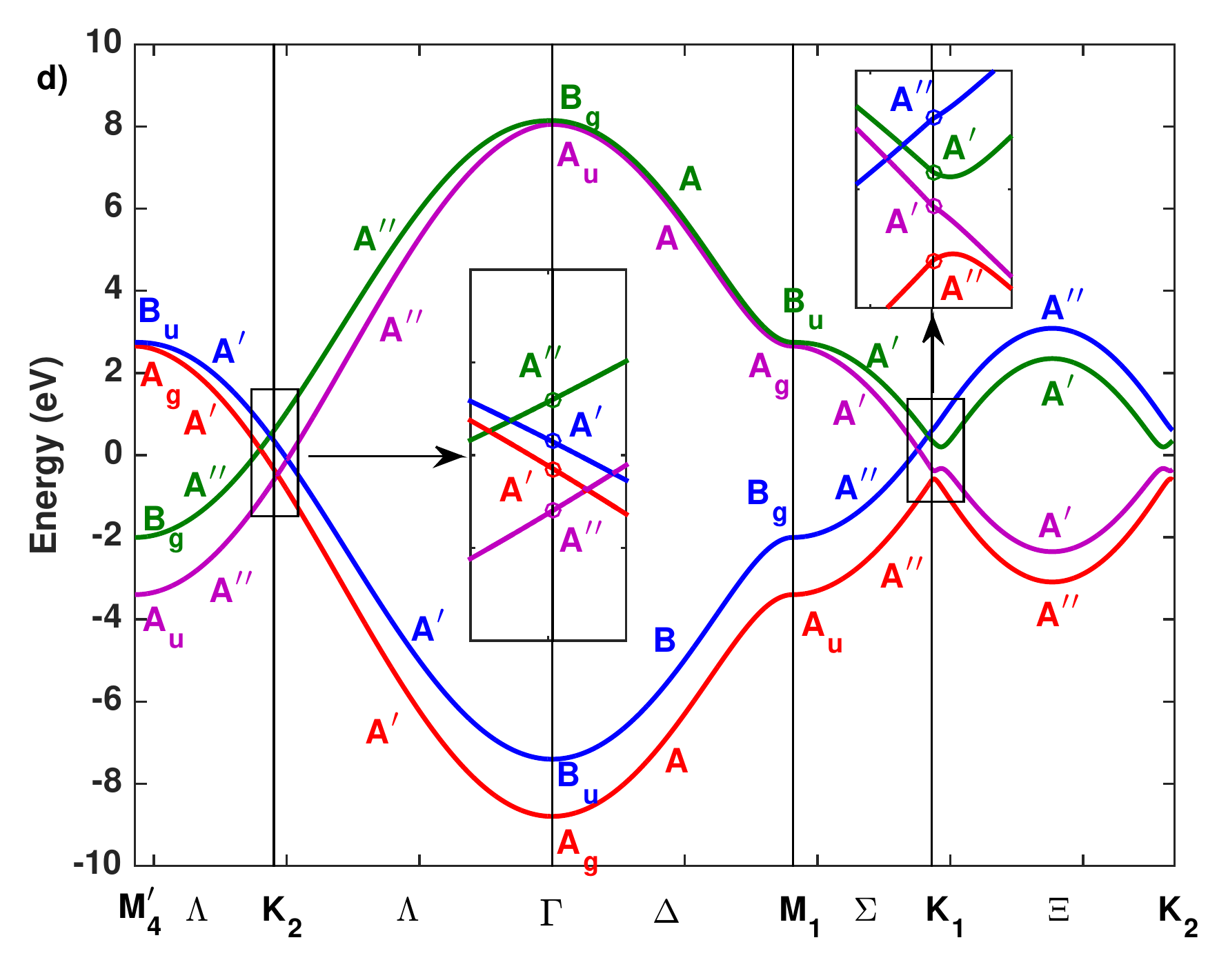}
\caption{\label{Fig2} Energy band structure of electrons in four typical SBG configurations: a) the $AA$-stacked, b) the $AB$-stacked, and the ones with $\boldsymbol{\tau}=\frac{\xi}{3}(\mathbf{a}_1+\mathbf{a}_2)$ (c) and $\boldsymbol{\tau}=\frac{\zeta}{2}(\mathbf{a}_1-\mathbf{a}_2)$ (d). All bands at the points and along the axes of high symmetry are labeled by the corresponding irreducible representations. The symmetry of each dispersion curve is distinguished by color.}
\end{figure*}

\begin{figure*}\centering
\includegraphics[clip=true,trim=7cm 8.5cm 6cm 8cm,width=0.2\textwidth]{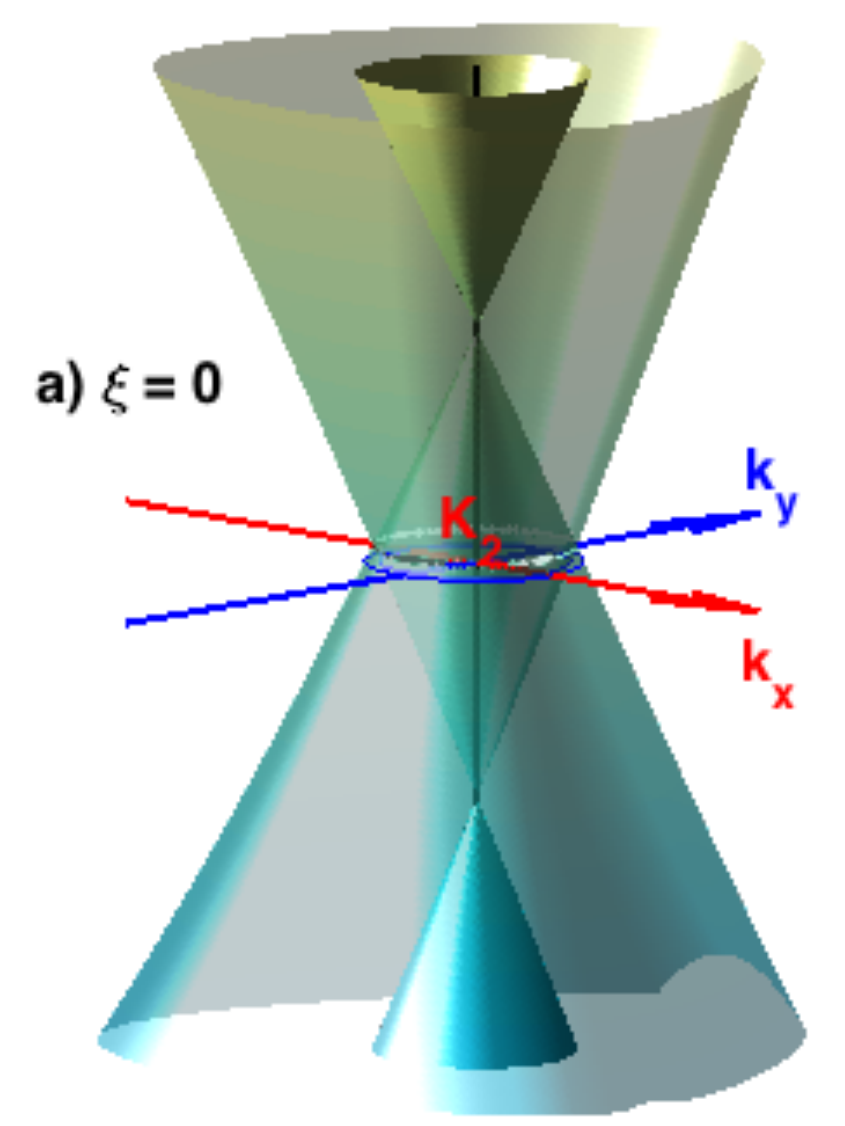}
\includegraphics[clip=true,trim=7cm 8.5cm 6cm 8cm,width=0.2\textwidth]{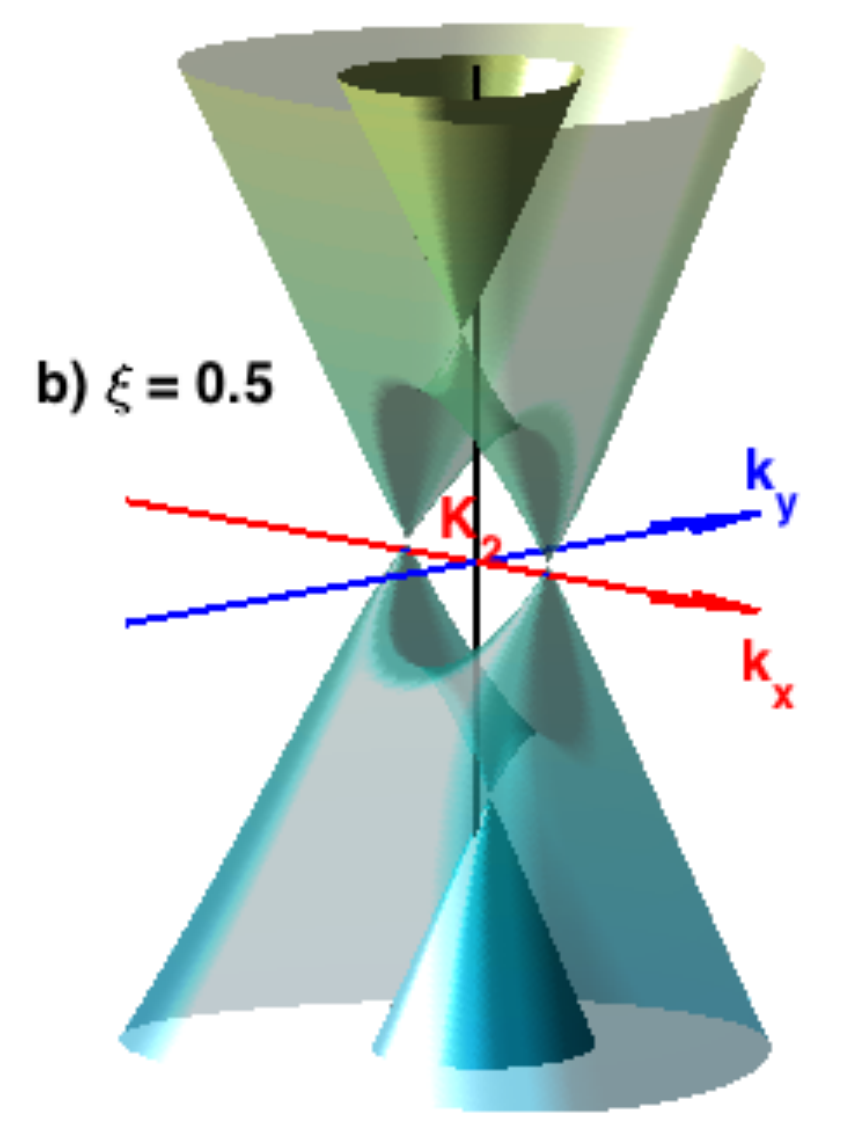}
\includegraphics[clip=true,trim=7cm 8.5cm 6cm 8cm,width=0.2\textwidth]{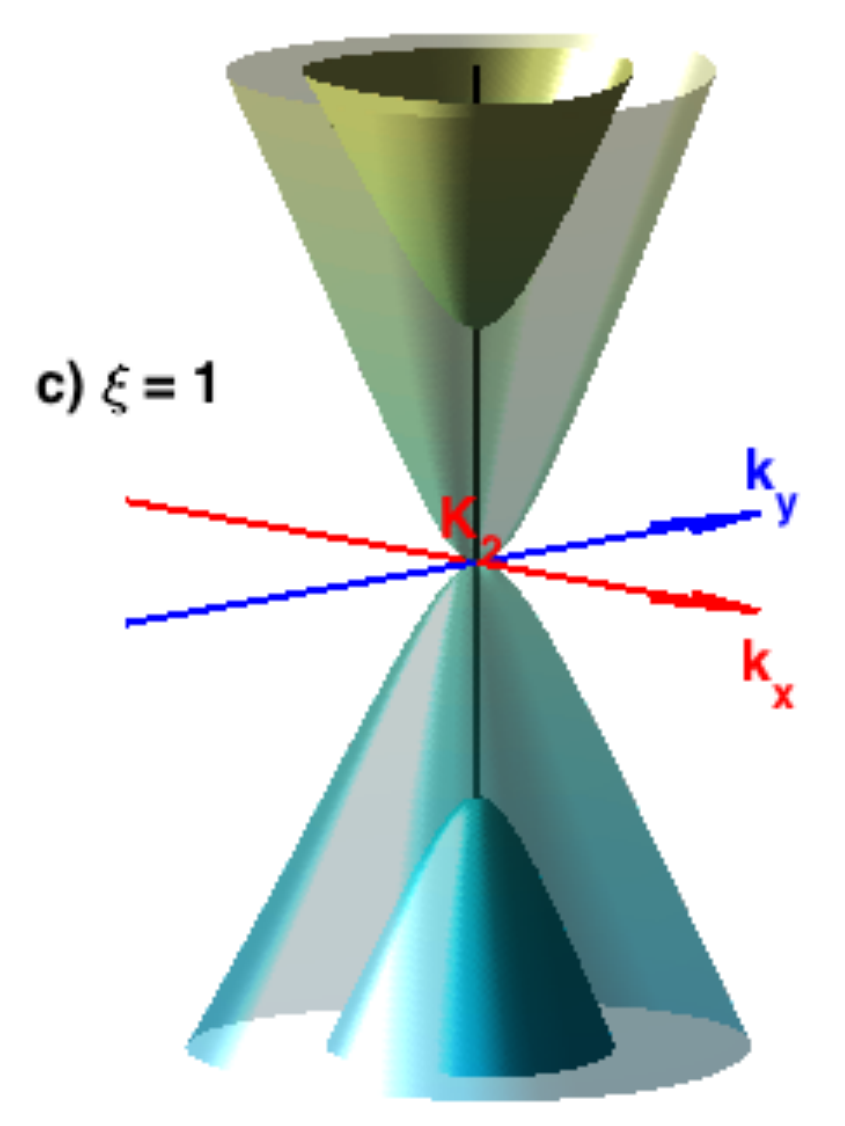}
\includegraphics[clip=true,trim=7cm 8.5cm 6cm 8cm,width=0.2\textwidth]{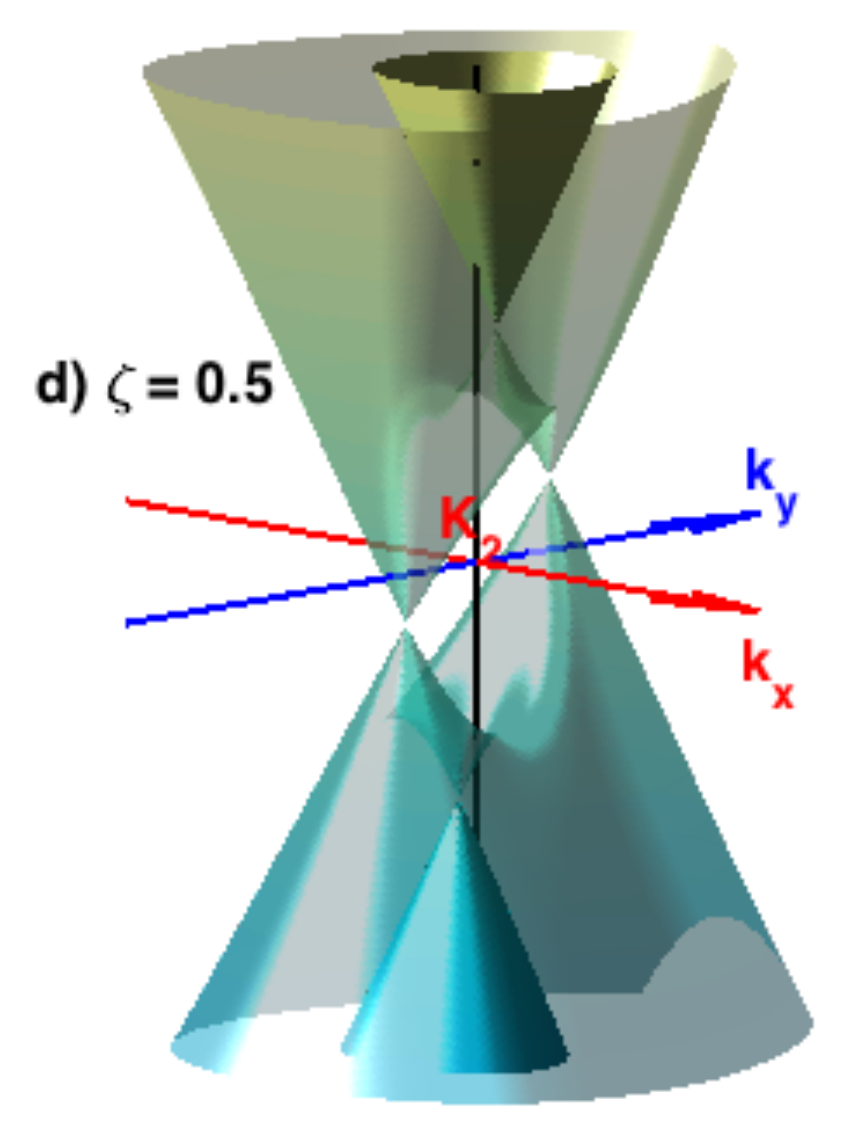}
\caption{\label{Fig3} Energy surfaces of four typical SBG configurations plotted around the $K_2$ point of the Brillouin zone.}
\end{figure*}

\subsection{Discussion of the electronic structure results}
Calculating the energy bands requires the diagonalization of the Bloch-Hamiltonian matrix. For the $4\times4$ matrix described in the previous section, see Eq. (\ref{Eq13}), the diagonalization generally needs to invoke numerical methods.  In Fig. \ref{Fig2} we present our calculated results for the energy dispersion curves for several typical SBG configurations along the high symmetry paths in the Brillouin zone. These are the $\Lambda$ path that connects the $M_4$, $K_2$ and $\Gamma$ points; the $\Delta$ path which connects the $\Gamma$ and $K_1$ points; and the $\Xi$ path that connects the $K_1$ and $K_2$ points (see the red lines in Fig. \ref{Fig1}(f)). We use colors to refer electronic states that are symmetrically compatible. Accordingly, we observe the overall picture of the smooth energy dispersion curves. By this way it is the proof for the crossing of the dispersion curves. We assign labels to the dispersion curves on the base of combining the analysis of the basis vectors of invariant subspaces (see Tables \ref{Table_V},\ref{Table_VI} and \ref{Table_VII}) and checking the symmetry of eigenstates of the Bloch-Hamiltonian matrix obtained by numerical calculations. From the obtained results one can observe that the lowest conduction energy surface $E_3(\mathbf{k})$ and the highest valence energy surface $E_2(\mathbf{k})$ are (near) degenerate at two $\mathbf{k}$ points in the vicinity of the $K$ points. In the high symmetry AA-stacked and AB-stacked configurations, the crossing of the dispersion curves occurs exactly at the $K$ points. The numerical results presented in Figs. \ref{Fig2}(a) and \ref{Fig2}(b) are compatible with the analysis of the representation of the symmetry group of the $K$ points in the 4-dimensional Hilbert space $\mathscr{H}=\text{span}(|A_1,p_z,\mathbf{k}\rangle,|B_1,p_z,\mathbf{k}\rangle,|A_2,p_z,\mathbf{k}\rangle,|B_2,p_z,\mathbf{k}\rangle)$. In particular, for the AA-stacked configuration, at the $K$ points, we see two crossing points involving four energy dispersion curves. Analysis of the representation of the $D_{3h}$ group at the $K$ points shows that the 4-dimensional Hilbert space $\mathscr{H}$ is separated into two invariant subspaces of 2-dimensions, $\mathscr{H}=\mathscr{H}(E^\prime)\oplus\mathscr{H}(E^{\prime\prime})$, (see Table V). The degeneracy of the energy at the crossing points is therefore equal to 2. For the AB-stacked configuration, we realize that the dispersion curves $E_2(\mathbf{k})$ and $E_3(\mathbf{k})$ touch each other at a single $K$ point, while the other two $E_1(\mathbf{k})$ and $E_4(\mathbf{k})$ curves separate. The analysis of the representation of the $D_3$ symmetry group for the $K$ points (see Table V) shows that the total Hilbert space can be divided into the direct sum of two 1-dimensional invariant subspaces and one 2-dimensional invariant subsspaces, $\mathscr{H}=\mathscr{H}(A_1)\oplus\mathscr{H}(A_2)\oplus\mathscr{H}(E)$. The vicinity of the crossing points in the energy dispersion curves of the AA- and AB-stacked configurations has been analyzed in literature.\cite{Rhokov_2016,McCann_2013} In particular, the dispersion curves for the AA-stacked configuration are linear in terms of $\delta\mathbf{k}=\mathbf{k-K}$, hence $E_n(\mathbf{k})\propto \|\delta\mathbf{k}\|$. Meanwhile, for the latter case, the dispersion curves display parabolic behavior near these points, i.e., $E_n(\mathbf{k})\propto \|\delta\mathbf{k}\|^2$. For the configurations with  $\boldsymbol{\tau}\propto \mathbf{a}_1\pm \mathbf{a}_2$, the dispersion curves cross the energy axis going through the $K$ point at 4 separate points. This result is also compatible with the group representation theory analysis. Accordingly, the 4-dimensional Hilbert space $\mathscr{H}$ representing the  $C_2$ or $C_s$ groups of the $K$ points is divided into the sum of four 1-dimensional invariant subspaces, $\mathscr{H}=2\mathscr{H}(A)\oplus 2\mathscr{H}(B)$, or  $\mathscr{H}=2\mathscr{H}(A^\prime)\oplus 2\mathscr{H}(A^{\prime\prime})$, respectively (see Tables \ref{Table_VI} and \ref{Table_VII}). Instead of coinciding at the $K$ points, the dispersion curves cross each other at the $\mathbf{k}$ points nearby the $K$ point. According to the result of symmetry group analysis shown in Table \ref{Table_IV} and in Fig. \ref{Fig2}, such crossing points of the dispersion curves are ensured by the compatibility relations between the symmetry of the $M,K$ and $\Gamma$ points.

\begin{figure}\centering
\includegraphics[clip=true,trim=0.3cm 6.5cm 0cm 7cm,width=\columnwidth]{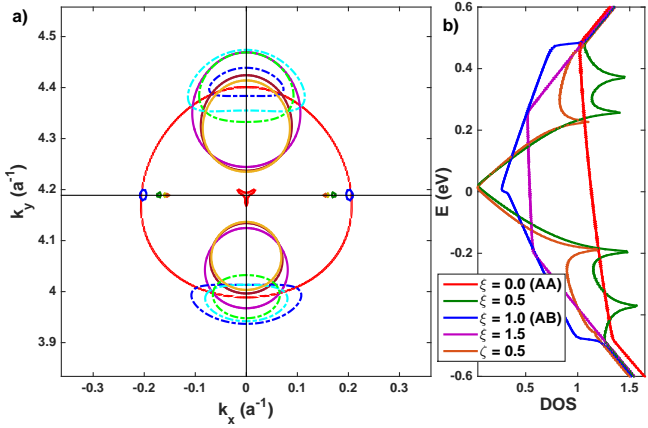}
\caption{\label{Fig4}(a) The Fermi energy surface of the AA-stacked (red large circle), AB-stacked (red small triangle at the centre) configurations. The tiny circles or dots on the $k_x$-direction are for the SBG configurations with $\xi = 0.05, 0.5$ and $0.8$. The pairs of solid circles are for the cases $\xi = 1.05, 1.2$ and 1.5. The pairs of dashed ellipse-like loops are for the cases of $\zeta = 0.05, 0.2$ and 0.5. (b) The density of states of five typical SBG configurations with $\xi = 0.0$ (red curve, the AA-stacked configuration), $\xi = 0.5$ (moss-green curve), $\xi = 1.0$ (blue curve, the AB-stacked configuration), $\xi = 1.5$ (purple curve), and $\zeta = 0.5$ (brow curve).}
\end{figure}

Analyzing the dispersion curves along the $\Lambda,\Delta, \Sigma$ and $\Xi$ paths would not be sufficient to fully understand  the behavior of the energy surfaces around the Fermi energy level. If one only looks at Fig. \ref{Fig2}(c) one may naively observe the opening of a finite narrow band gap. However, this picture would not be correct since there is no band gap in the electronic structure of the SBG configurations. In Fig. \ref{Fig3} we present the energy surfaces around the $K_2$ point as overall 3D view. From our calculations we can make the following observations:  (1) The electronic structure of the AA-stacked configuration is formed by the merging of the energy surfaces of two individual graphene layers. The merging manifests itself as a homogeneous potential of $-t_\perp$ for one layer and of $+t_\perp$ for the other. As a consequence, the highest valence surface and the lowest conduction surface cross each other via a circle on the zero-energy plane. This circle therefore defines the Fermi energy surface. (2) Sliding two graphene layers past each other (i.e., $\boldsymbol{\tau}\neq 0$) leads to the relative shift of the energy surfaces corresponding to each graphene layer along the $Oy$ direction (i.e., along the symmetry axis that is free from the constraints imposed by the lattice symmetries). (3) The shift of the energy surfaces occurs together with the deformation of energy surfaces in the energy range of  $(-t_\perp,+t_\perp)$ due to the hybridization of the electronic states of the individual graphene layers. (4) The deformation of the energy surfaces is not able to open a finite band gap but forms a structure of the highest valence and lowest conduction surfaces touching each other at two Dirac points around the $K$ point. In the case of $\boldsymbol{\tau} = \xi(\mathbf{a}_1+\mathbf{a}_2)/3$ where $\xi \in (0,1)$, the two Dirac points lie on the $Ox$ direction on the same energy plane $E=0$. The Fermi energy surface is therefore determined as the set of points around the $K$ points (see Fig. \ref{Fig4}(a)). Meanwhile, for $\boldsymbol{\tau} = \xi(\mathbf{a}_1+\mathbf{a}_2)/3$ where $\xi \in (1,3/2]$, and $\boldsymbol{\tau} = \zeta(\mathbf{a}_1-\mathbf{a}_2)/2$, the two Dirac points lie along the $Oy$ direction but not on the same energy plane; one is shifted upward while another is shifted downward to form the structure of two tilted miniature cones. The Fermi energy surface therefore takes the form of two separated circles centered along the $Oy$ direction. (5) For the SBG configurations that are far from the AB-stacked configuration, the two Dirac points of each graphene layer (one located on the $E=+t_\perp$ energy plane and the other on the $E=-t_\perp$ energy plane) are preserved because the energy surfaces are strongly deformed only in the lower energy range around the Fermi level $E=0$. These two Dirac points move from the $K$ points because of the shift in the energy surfaces. In contrast, for the AB-stacked configuration, the hybridization of the electronic states in the two graphene layers is not the same between two sub-lattices. It therefore causes the destruction of the two Dirac points and parabolically separates the $E_{4/1}(\mathbf{k})$ surfaces from the $E_{3/2}(\mathbf{k})$ surfaces. 

In order to elucidate further the formation of the Dirac points in the electronic structure of the SBG configurations it is instructive to carefully revisit the highly symmetric  AA-stacked configuration. As described above, the energy  calculations show that around the $K$ points, the electronic structure is formed as the merging of the energy surfaces of two individual graphene layers, each one is shifted by a certain amount of energy. This result is generally explained in literature as a consequence of the chiral symmetry of the dynamical model.\cite{Hatsugai_2013} With this in mind, for this configuration it is sufficient to consider the dominant interlayer coupling between the $p_z$ orbitals located at the nodes $A_1$ and $A_2$, as well as the nodes $B_1$ and $B_2$. Indeed, we have a quantitative comparison of the magnitude of the hopping parameters as follows: $t_0 = 2.7 \,\text{eV} \gg t_{A_1A_2}=t_{B_1B_2}=t(\mathbf{d}_z)=0.48 \,\text{eV} \gg t_{A_1B_2} = t_{B_1A_2}=t(\mathbf{d}_z+\mathbf{d}_1) = 0.18 \,\text{eV}$. Therefore, the Bloch-Hamiltonian matrix elements involving in the terms $v_\mathbf{k}$ and $w_\mathbf{k}$ can be approximately set to zero. The only nonzero matrix elements besides the $f_\mathbf{k}$ elements are the $u_\mathbf{k}$ elements. They are the most dominant among the interlayer terms. The approximate Bloch-Hamiltonian becomes:
\begin{equation}\label{Eq16}
    h_{AA}(\mathbf{k}) = \left(\begin{array}{cccc}
    0&f_\mathbf{k}& u_\mathbf{k}&0\\
    f^*_\mathbf{k}&0&0&u_\mathbf{k}\\
    u^*_\mathbf{k}&0&0&f_\mathbf{k}\\
    0&u^*_\mathbf{k}&f^*_\mathbf{k}&0
    \end{array}\right).
\end{equation}
An important consequence of the approximations is that  chiral symmetry is introduced into the Bloch-Hamiltonian matrix. Indeed, by rearranging the order of the basis vectors representing the Bloch-Hamiltonian matrix into $\{|A_1,\mathbf{k}\rangle,|B_2,\mathbf{k}\rangle;|A_2,\mathbf{k}\rangle,|B_1,\mathbf{k}\rangle\}$, we get:
\begin{equation}\label{Eq17}
    h_{AA}(\mathbf{k}) = \left(\begin{array}{cccc}
    0&0&u_\mathbf{k}&f_\mathbf{k}\\
    0&0&f^*_\mathbf{k}&u^*_\mathbf{k}\\
    u^*_\mathbf{k}&f_\mathbf{k}&0&0\\
    f^*_\mathbf{k}&u_\mathbf{k}&0&0
    \end{array}\right).
\end{equation}
Chiral symmetry implies that the eigenvalues of the matrix $H(\mathbf{k})$ will always appear in pairs of $\pm E(\mathbf{k})$, and if zero-energy states exist, they must be degenerate. We will first  consider the existence of the zero-energy states. The eigenstates corresponding to  $E=0$ are defined by the vectors whose coordinates (in the set of basis vectors under consideration) must satisfy the homogeneous linear equations:
\begin{equation}\label{Eq18}
    \left(\begin{array}{cccc}
    0&0&u_\mathbf{k}&f_\mathbf{k}\\
    0&0&f^*_\mathbf{k}&u^*_\mathbf{k}\\
    u^*_\mathbf{k}&f_\mathbf{k}&0&0\\
    f^*_\mathbf{k}&u_\mathbf{k}&0&0
    \end{array}\right)\left(\begin{array}{c}x\\y\\z\\s\end{array}\right)=\left(\begin{array}{c}0\\0\\0\\0\end{array}\right).
\end{equation}
After some algebraic manipulation, we obtain an equation which defines the condition for the existence of the zero-energy states:
\begin{equation}\label{Eq19}
    f^*_\mathbf{k}f_\mathbf{k}-u^*_\mathbf{k}u_\mathbf{k}=0.
\end{equation}
This equation actually determines the value of the vector $\mathbf{k}$ defining the corresponding zero-energy states. It is straightforward to deduce:
\begin{equation}\label{Eq20}
    z = -\frac{f_\mathbf{k}}{u_\mathbf{k}}s; y = -\frac{f_\mathbf{k}^*}{u_\mathbf{k}}x.
\end{equation}
The zero-energy state vectors are determined by:
\begin{equation}\label{Eq21}
    |E=0,\mathbf{k}\rangle = x|E=0,\mathbf{k}\rangle_1+y|E=0,\mathbf{k}\rangle_2,
\end{equation}
with
\begin{subequations}
\begin{align}
    |E=0,\mathbf{k}\rangle_1 &=\frac{1}{\sqrt{2}}\left( |A_1,\mathbf{k}\rangle-\frac{f_\mathbf{k}^*}{u_\mathbf{k}}|B_2,\mathbf{k}\rangle\right)\label{Eq22a},\\
    |E=0,\mathbf{k}\rangle_2 &=\frac{1}{\sqrt{2}}\left( |B_1,\mathbf{k}\rangle-\frac{f_\mathbf{k}}{u_\mathbf{k}}|A_2,\mathbf{k}\rangle\right)\label{Eq22b}.
\end{align}
\end{subequations}
From the equations it is clear that the zero-energy states are two-fold degenerate, as expected. Theses states belong to a 2-dimensional Hilbert space spanned by two orthogonal state vectors given by Eqs. (\ref{Eq22a}) and (\ref{Eq22b}). Remarkably, these states are the linear combination of the states locally defined in two sub-lattices that are not directly coupled with each other. Next we will show that Eq. (\ref{Eq19}) defines a the Fermi energy surface as the set of $\mathbf{k}$ points in a circle centered at the $K$ point and with the radius of $t_\perp/\hbar v_F$.

Given the chiral symmetry structure of the Bloch-Hamiltonian matrix of Eq. (\ref{Eq17}) it is straightforward to deduce an analytic solution for the eigenvalues $E_n(\mathbf{k})$ of  $h_{AA}(\mathbf{k})$. Indeed, using the rule for the  determinant of the block matrices (see Appendix A) the secular equation for the nonzero eigenvalues is written as:
\begin{equation}\label{Eq23}
    \det\left(\begin{array}{cc}
    f^*_\mathbf{k}f_\mathbf{k}+u^*_\mathbf{k}u_\mathbf{k}-E^2   &2f_\mathbf{k}u_\mathbf{k}  \\
    2f^*_\mathbf{k}u^*_\mathbf{k}&f^*_\mathbf{k}f_\mathbf{k}+u^*_\mathbf{k}u_\mathbf{k}-E^2 
    \end{array}\right) = 0.
\end{equation}
The four eigenvalues are:
\begin{subequations}
\begin{align}
    E_{1,2}(\mathbf{k})&=+\sqrt{u^*_\mathbf{k}u_\mathbf{k}}\pm\sqrt{f^*_\mathbf{k}f_\mathbf{k}},\label{Eq24a} \\
    E_{3,4}(\mathbf{k})&=-\sqrt{u^*_\mathbf{k}u_\mathbf{k}}\pm\sqrt{f^*_\mathbf{k}f_\mathbf{k}}.\label{Eq24b}
\end{align}
\end{subequations}
Noting that at the $K$ points we have $f_\mathbf{K}=0$ and $u_\mathbf{K} = t_\perp$, in the vicinity of the $K$ points we therefore have the expansion:
\begin{equation}\label{Eq25}
    f_{\mathbf{K+\delta k}} \approx \delta\mathbf{k}\cdot \left(\partial_\mathbf{k}f_\mathbf{k}\right)_{\mathbf{K}}
\end{equation}
Using Eq. (\ref{Eq15}) for the expression of $f_\mathbf{k}$ results in 
\begin{eqnarray}\label{Eq26}
    \left(\partial_\mathbf{k}f_\mathbf{k}\right)_{\mathbf{K}} = i\frac{3}{2}t_0a_{CC}e^{i\frac{2\pi}{3}}(1,i)\equiv ie^{i\frac{2\pi}{3}}\hbar v_F(1,i).
\end{eqnarray}
We thus obtain:
\begin{equation}\label{Eq27}
    \sqrt{f^*_\mathbf{k}f_\mathbf{k}}\approx |\delta\mathbf{k}\cdot\left(\partial_\mathbf{k}f_\mathbf{k}\right)_\mathbf{K}| = \hbar v_F\|\mathbf{\delta k}\|.
\end{equation}
Hence we can see that the energy surfaces around the $K$ points of the AA-stacked configuration are cones.\cite{McCann_2013,Rhokov_2016} In particular, the two surfaces $E_1(\mathbf{k})$ and $E_2(\mathbf{k})$ can be seen as the energy surfaces of one graphene layer that is subjected to a potential equal to $u_\mathbf{K} = t_\perp$. Similarly, the surfaces $E_3(\mathbf{k})$ and $E_4(\mathbf{k})$ are the ones consisting of the electronic structure of the second graphene layer subjected to a potential of $-t_\perp$. Consequently, the two energy surfaces $E_{2}(\mathbf{k})$ and $E_3(\mathbf{k})$ cross each other at the zero energy level. The set of  $\mathbf{k}$ points lying on the crossing defined by the equation $\|\delta\mathbf{k}\| =t_\perp/\hbar v_F$. This is the equation for a circle with its center located at the $K$ points and the radius of $t_\perp/\hbar v_F$. This equation defines the Fermi energy surface of the AA-stacked configuration.

The use of a model for the chiral symmetry was already presented in literature.\cite{McCann_2013,Park_2015,Rhokov_2016,Hatsugai_2013} It is, however, much more intuitive to understand why the electronic structure of the AA-stacked configuration can be simply seen as the merging of the energy surfaces of the two individual graphene layers by noting the mirror symmetry $M_{xy}$. Hence by using the $U(\mathbf{k})=(\tau_z+\tau_x)\otimes\sigma_0/\sqrt{2}$ transformation (where $\tau_x,\tau_z$ are the first and third Pauli matrices, respectively, and $\sigma_0$ is the $2\times 2$ identity matrix) the current Bloch basis set  $\{|\alpha,p_z,\mathbf{k}\rangle\}$ can be transformed so that the Bloch-Hamiltonian matrix is a diagonal block matrix. In this form, each $2\times 2$ diagonal block is exactly the Bloch-Hamiltonian matrix of a single graphene subjected to the potential of $-u_\mathbf{k}$ or $u_\mathbf{k}$: 
\begin{equation}\label{Eq28}
    h_{AA}(\mathbf{k}) = \left(\begin{array}{cccc}
    u_\mathbf{k}&f_\mathbf{k}&0&0\\
    f^*_\mathbf{k}&u_\mathbf{k}&0&0\\
    0&0&-u^*_\mathbf{k}&f_\mathbf{k}\\
    0&0&f^*_\mathbf{k}&-u^*_\mathbf{k}
    \end{array}\right).
\end{equation}
The physics of this result can be understood in terms of the electronic interlayer coupling between  the two graphene layers in the high symmetry $AA$-stacked configuration not breaking the symmetry between the two sub-lattices $A$ and $B$ of the individual graphene layers. As a consequence, an electron in one layer simply sees the presence of the second layer through a homogeneous and isotropic potential. 

Significant sliding between the two graphene layers  dramatically breaks the symmetry of the system. In particular, the rotational symmetries around the axes perpendicular to the lattice plan will disappear and thus the initial symmetry group $P6/mmm$ is lowered to $P2/m$. In such cases, it is necessary to take into account the matrix elements $v_\mathbf{k}$ and $w_\mathbf{k}$ since they play an important role in the mixing of the electronic states between the graphene layers. Consequently, the energy surfaces in the range from $-t_\perp$ to $+t_\perp$ are strongly deformed. Though the degeneracy is lifted up at the crossing loop between the energy surfaces $E_2(\mathbf{k})$ and $E_3(\mathbf{k})$, a finite energy gap is not fully created. In particular, for the cases where $\boldsymbol{\tau} = \xi(\mathbf{a}_1+\mathbf{a}_2)/3$ with $\xi\in(0,1)$, the energy surfaces $E_{1,2}(\mathbf{k})$ and $E_{3,4}(\mathbf{k})$ are oppositely shifted along the $Oy$ direction. Meanwhile, in the same energy range the two surfaces $E_2(\mathbf{k})$ and $E_3(\mathbf{k})$ shrink to form a structure of two mini cones touching each other at their vertices. The two vertices are located at two opposite $\mathbf{k}$ points on the circle $\|\delta\mathbf{k}\|=t_\perp/\hbar v_F$, along the $Ox$ direction. Around these two vertices, despite the dispersion relation being linear, the cones are anisotropic. The issue of two Dirac points lying on the $E=0$ plane can be explained as the preservation of a symmetry operation under the sliding of two graphene layers. Indeed, we see that when the $D_{6h}$ group is broken to the $C_{2h}$ group, the symmetry group of the $K$ points reduces from $D_{3h}$ to $C_2$. In this case, the two-fold $C_{2y}$ symmetry is preserved. This symmetry imposes a constraint on the touching of the energy surfaces  $E_2(\mathbf{k})$ and $E_3(\mathbf{k})$ at two $\mathbf{k}$ points on the circle $\|\delta\mathbf{k}\|=t_\perp/\hbar v_F$ along the $Ox$ direction. When the sliding vector is long enough, $\xi\rightarrow 1$, such that the SBG configuration is transformed into the AB-stacked configuration, the two mini Dirac points can move out of the circle $\|\delta\mathbf{k}\|=t_\perp/\hbar v_F$. They approach each other and then merge together at the $K$ point when the group $C_{2h}$ becomes $D_{3d}$. In the cases that $\xi\in (1,3/2]$, the electronic structure of the SBG configuration can be seen as the result of the deformation of the electronic structure of the AB-stacked configuration ($\xi = 1$). In the latter case, the four energy surfaces have the form of the isotropic parabolic surfaces aligned in the same axis. When $\xi > 1$, the opposite shift of the surfaces $E_{1,2}(\mathbf{k})$ and $E_{3,4}(\mathbf{k})$ occurs along the $Oy$ direction because of the symmetry breaking, from the group  $P\bar{3}m1$ to $P2/m$ (or from the point group $D_3$ to  $C_2$ for the $K$ points). Consequently, it leads to the crossing of the dispersion curves $E_{1,2,3,4}(0,k_y)$ as shown in Fig. \ref{Fig2}(d). At the same time, sliding along the $Ox$ direction causes the deformation of the $E_2(\mathbf{k})$ and $E_3(\mathbf{k})$ surfaces along the $Ox$ direction to result in the formation of the two mini cones with their vertices not lying on the same energy plane. According to the previously shown analysis of the symmetry compatibility relations among the points $M_4, K_2$ and $\Gamma$, the re-emergence of the Dirac points in the SBG electronic structure is guaranteed by symmetry. In the cases that the sliding vector $\tau$ is directed along the $Oy$ direction, $\boldsymbol{\tau} = \zeta(\mathbf{a}_1-\mathbf{a}_2)/2, \zeta\in (0,1]$, the shifting of the graphene layer energy surfaces takes place along the $Oy$ direction (the axis that is free of the constraints imposed by lattice symmetries), leading to a picture similar to that of the case $\boldsymbol{\tau} = \xi(\mathbf{a}_1+\mathbf{a}_2)/3, \xi>1$. However, one should notice that the symmetry operation $M_{yz}$ is preserved in the former case instead of the two-fold rotation operation $C_{2y}$ in the latter case. Though these two symmetry operations cause the same effect on $\mathbf{k}$, i.e., transforming $(k_x,k_y)$ into $(-k_x,k_y)$, they impose quantitatively different constraints on the energy surfaces, i.e., $E_n(k_x,K_{2y})=E_n(-k_x,K_{2y})$ for the former and  $E_{2,1}(k_x,K_{2y}) = E_{3,4}(-k_x,K_{2y})$ for the latter.

Finally, we present in Fig. \ref{Fig4} the Fermi energy surface and the electronic density of states (DOS) for  several SBG configurations. In Fig. \ref{Fig4}(a), the large red circle is the part of the Fermi energy surface around the $K$ point of the AA-stacked configuration. As discussed above, it is the locus of the $\mathbf{k}$-points defining the zero-energy surface. This closed path, described by Eq. (\ref{Eq19}), is identical to the crossing of the two cone surfaces $E_2(\mathbf{k})$ and $E_3(\mathbf{k})$. When $\xi = 0.05$, though small, sliding of the bilayers reduces the symmetry of the AA-stacked configuration from $P6/mmm$ to $P2/m$. As a consequence, the Fermi surface collapses onto the two (blue) points on the red circle as shown in the figure. For large values of $\xi$, these two points go off the red circle and approach each other towards the position of the $K$ point. As discussed, it is the result of the shift and shrinking of the energy surfaces in the energy range of $(-t_\perp,+t_\perp)$ during the formation of the Dirac points. For the cases where $\xi > 1.0$ and $\zeta > 0$, the Fermi surface around the $K$ points takes the shape of two separate circles centered on the $k_y$ axis (the solid curves for the former cases and the dashed curves for the latter ones). It reflects the (anisotropic) titling cone structure of the energy surfaces. Fig. \ref{Fig4}(a) provides a picture of the topological transition of the Fermi energy surface with the sliding two graphene layers. This so-called Lifshitz transition was also discussed in Refs. \onlinecite{Son_2011,Bhattacharyya_2016,Suszalski_2018,Jayaraman_2021} for small value of the sliding vector. The shape of the energy surfaces and the topological structure of the Fermi surface reflects in the picture of the density of states. The red and blue curves of Fig. \ref{Fig4}(b) are the DOS of the AA- and AB-stacked configurations and they have been thoroughly studied in literature.\cite{McCann_2013,Rhokov_2016} In the same figure, the moss-green, purple and brown curves are for the SBG configurations with $\xi = 0.5, 1.5$ and $\zeta = 0.5$, respectively. While the purple curve takes the form of the red one with a constant value in a narrow energy range, the dark-green and brown curves show the typical V-shape, that is characteristic of the linear dispersion law, for the cone surfaces of the monolayer graphene. The significant peaks of the curves are the manifestations of the saddle points in the energy surfaces. These results are consistent with the behavior of the energy surfaces shown in Fig. \ref{Fig3}.

\section{Conclusion}
The engineering of stacked layered materials to have desired electronic properties is currently an intensely developed field. One such actively researched materials is bilayer graphene, which is a flexible 2D system consisting of two graphene monolayers that are weakly bound together. We  investigated the electronic structure features of a special class of bilayer graphene configurations: the sliding bilayer graphene systems. We systematically studied the geometrical and topological properties of the energy surfaces of SBG configurations while varying the values of the sliding vector. Using a tight-binding model that takes into account only the atomic $p_z$ orbitals, we showed that the electronic structure of SBGs is formed by the merging of the energy surfaces of two individual graphene layers. Sliding two graphene layers causes two things. First, sliding along an axis that is free of constraints due to the lattice symmetries only shifts the energy surfaces of the individual graphene layers. Second, sliding of the bilayers breaks the symmetry in the interlayer coupling of the electronic states for the two graphene layers. Both result in the crossing of the dispersion curves along the symmetrical axes and the shrinking of the energy surfaces in a narrow range of energies around the Fermi level. The emergence of the Dirac points in the vicinity of the $K$ points is shown to be the result of the deformation of the energy surfaces under the constraints of the existing symmetries. These observations were validated using  analysis of the group representation theory. We prove that the band crossings at generic $\mathbf{k}$ points are guaranteed by the compatibility relations between the symmetries of the eigenstates at the high symmetry points in the Brillouin zone. The emergence of Dirac points define the geometrical and topological features of the energy surfaces, i.e., the local maximal, minimal and saddle points. They manifest in the electronic properties through the shape of the Fermi energy surface and the density of states.

\section*{Acknowledgements}
The author ackowledges Dominik Domin, Dario Bercioux and Miguel \'Angel Jim\'enez Herrera  for fruitful discussions and, especially, reading carefully the manuscript before it was submitted. 

\appendix
\section{Determinant of block matrices}
In general, let us consider the following block matrix:
\begin{equation}\label{A1}
    \left(\begin{array}{cc}
    A_{NN} & B_{NM}  \\
    C_{MN} & D_{MM} 
    \end{array}\right),
\end{equation}
where $M,N$ are the sizes of the submatrices (blocks). If the block $D_{MM}$ is invertible, the original block matrix can be transformed into a triangular matrix by the following matrix multiplication:
\begin{widetext}
\begin{equation}\label{A2}
    \left(\begin{array}{cc}
    A_{NN} & B_{NM}  \\
    C_{MN} & D_{MM} 
    \end{array}\right)\left(\begin{array}{cc}
    I_{NN} & 0_{NM}  \\
    -D^{-1}_{MM}C_{MN} & I_{MM} 
    \end{array}\right) =
    \left(\begin{array}{cc}
    A_{NN}-B_{NM}D_{MM}^{-1}C_{MN} & B_{NM}  \\
    0_{MN} & D_{MM} 
    \end{array}\right).
\end{equation}
\end{widetext}
The definition of the determinant allows one to deduce the following expression for the determinant of triangular block matrices:
\begin{equation}\label{A3}
    \det\left(\begin{array}{cc}
    A_{NN} & 0_{NM}  \\
    C_{MN} & D_{MM} 
    \end{array}\right)=\det(A_{NN})\det(D_{MM}).
\end{equation}
The determinant of the second matrix in Eq. (\ref{A2}) is trivially equal to one, hence one can combine Eqs.(\ref{A2}) and (\ref{A3}) to get the following expression:
\begin{align}\label{A4}
    \det\left(\begin{array}{cc}
    A_{NN} & B_{NM}  \\
    C_{MN} & D_{MM} 
    \end{array}\right) =& \det(A_{NN}-B_{NM}D_{MM}^{-1}C_{MN})\nonumber\\
    &\times\det(D_{MM}).
\end{align}
Applying this formula to Hamiltonian block matrices, one can readily obtain the secular equations seen in Eq. (\ref{Eq23}).

\section{Chiral symmetry}
An electronic system is said to possess chiral symmetry if  it is decomposible into subsystems  and there exists a local unitary Hermitian operator, $C$, that anticommutative with the Hamiltonian of the system. This means that $C H(\lambda) = - H(\lambda)C$, where $\lambda$ refers to a set of parameters which define the Hamiltonian $H$, but not $C$ (i.e., $C$ is independent of $\lambda$). These basic properties of the $C$ operator allow one to write:
\begin{subequations}
\begin{align}
    & C^2 = 1,\label{B1a}\\
    & C = \sum_n C_n,\label{B1b}\\
    & C H(\lambda)C = -H(\lambda).\label{B1c}
\end{align}
\end{subequations}
Here Eq. (\ref{B1a}) expresses the unitary and Hermitian properties of the chiral operator; the locality is expressed by Eq. (\ref{B1b}) as the decomposition into a set of unitary operators $C_n$ that act only on subsystem $n$. 

The hexagonal lattice of monolayer graphene can be seen as the composed of two sublattices $A$ and $B$. A chiral operator $C$ can be defined in terms of projection operators ($P_A$ and $P_B$) in the $A$ and $B$ sublattices:
\begin{equation}
    C = P_A-P_B,\label{B2}
\end{equation}
where $P_A+P_B = 1$ and $P_AP_B = 0$. The system's Hamiltonian $H$ is therefore decomposable into four terms describing the $A$ and $B$ sublattices and the coupling between them, i.e., $H = H_{AA}+H_{BB}+H_{AB}+H_{BA}$. The terms are defined as $H_{AA} = P_AHP_A, H_{BB}=P_BHP_B, H_{AB} = P_AHP_B$ and $H_{BA}=P_BHP_A$. If the system possesses chiral symmetry, from Eq. (\ref{B1c}) it can be deduced that the terms $H_{AA}$ and $H_{BB}$ must vanish. This is the case for a commonly used tight-binding Bloch-Hamiltonian matrix for non-interacting electrons in the nearest neighbor approximation:
\begin{equation}\label{B3}
    h_{MLG}(\mathbf{k}) = \left(\begin{array}{cc}
        0 & f_\mathbf{k} \\
        f^*_\mathbf{k} & 0
    \end{array}\right).
\end{equation}
Sliding bilayer graphene systems do not always possess  chiral symmetry. In high symmetry configurations, such as AA- and AB-stacked, some hopping terms between the four sublattices, $A_1,B_1,A_2$ and $B_2$, can be approximately ignored since they are much smaller than other terms. This may result in a Bloch-Hamiltonian matrix possessing the chiral symmetry. In particular, the Bloch-Hamiltonian matrix of the AA-stacked configuration given by Eq. (\ref{Eq17}) takes the form of off-diagonal block matrices. Chiral symmetry is present in this model. The bilayer lattice is decomposable into two sub-systems, one consisting of the sub-lattices $A_1$ and $B_2$ and the other of the sub-lattices $B_1$ and $A_2$. The zero-energy eigenstates of $h_{AA}(\mathbf{k})$ given by Eqs. (\ref{Eq22a}) and (\ref{Eq22b}) are clearly the linear combinations of only two atomic orbitals $A_1$ and $B_2$ or $B_1$ and $A_2$. These results reflect the localization of the zero-energy eigenstates on either the lattice nodes $(A_1,B_2)$ or on the lattice nodes $(B_1,A_2)$.

\section{Symmetry analysis}
In the main text, $T_g$ is a unitary linear operator representing a symmetry operation $g$ of a symmetry group $G(\mathbf{k})$ for a $\mathbf{k}$ vector within the Brillouin zone. The operators $T_g,\,\forall\, g\in G(\mathbf{k})$, act on the Hilbert space spanned by four orthogonal Bloch vectors $\{|A_1,p_z,\mathbf{k}\rangle,|B_1,p_z,\mathbf{k}\rangle,|A_2,p_z,\mathbf{k}\rangle,|B_2,p_z,\mathbf{k}\rangle\}$. Using Eq. (\ref{Eq6}) we can represent $T_g$ operators in terms of matrices as follows:
\begin{equation}\label{C1}
    T_g|\alpha,p_z,\mathbf{k}\rangle = \sum_{\alpha^\prime}s_g\widetilde{U}^g_{\alpha^\prime\alpha}|\alpha^\prime,p_z,g\mathbf{k}\rangle,
\end{equation}
where $\alpha^\prime$ is the result of the transformation of the sub-lattice index $\alpha$ under the operation of $g$, i.e., $\alpha^\prime = g\alpha$. The other terms in the expression are the sign $s_g$ of the orbital $p_z$ under the operation $g$ and the coefficients, $\widetilde{U}^g_{\alpha^\prime\alpha}$, for a particular linear combination of the Bloch vectors. Since $\mathbf{k}$ is a symmetrical point in the Brillouin zone, the action of the symmetry operation $g\in G(\mathbf{k})$ results in another point $\mathbf{k}^\prime = g\mathbf{k}$ that belongs to the star of $\mathbf{k}$. This means that $g\mathbf{k} - \mathbf{k} = \mathbf{G}_g$ is a reciprocal lattice vector. Now, using Eq. (\ref{Eq5}) for the definition of the Bloch vector and adding the reciprocal lattice vector to the $\mathbf{k}$-vector:
\begin{equation}\label{C2}
    |\alpha^\prime,p_z,\mathbf{k}+\mathbf{G}_g\rangle = e^{-i\mathbf{G}_g\cdot\mathbf{d}_{\alpha^\prime}}|\alpha^\prime,p_z,\mathbf{k}\rangle.
\end{equation}
Th elements of the $U_g(\mathbf{k})$ matrix that represens the $T_g$ operator can be obtained by inserting the basis vector from Eq. (\ref{C2}) into Eq. (\ref{C1}) and simplifying the expression to:
\begin{equation}\label{C4}
    T_g|\alpha,p_z,\mathbf{k}\rangle = \sum_{\alpha^\prime}U^g_{\alpha^\prime\alpha}(\mathbf{k})|\alpha^\prime,p_z,\mathbf{k}\rangle,
\end{equation}
where
\begin{equation}\label{C5}
    U^g_{\alpha^\prime\alpha}(\mathbf{k}) = s_g\widetilde{U}^g_{\alpha^\prime\alpha}e^{-i\mathbf{G}_g\cdot\mathbf{d}_{\alpha^\prime}}.
\end{equation}

Since $G(\mathbf{k})$ is a sub-group of a symmetry group of the system's Hamiltonian $H$, the invariant relation $T_gHT_g^{\dagger} = H$ allows one to deduce an identical relation for the Bloch-Hamiltonian matrix $h(\mathbf{k})$:
\begin{equation}\label{C6}
    h(\mathbf{k}) = U_g(\mathbf{k})h(\mathbf{k})U_g^\dagger(\mathbf{k}).
\end{equation}
This equation implies that the Bloch-Hamiltonian matrix is invariant under the transformation matrix $U_g(\mathbf{k})$. We therefore check the symmetry of the electronic bands using the following procedure. For every symmetrical $\mathbf{k}$ point in the Brillouin zone we: (1) find the $U_g(\mathbf{k})$ matrices that represent the symmetry operations of the small symmetrical group $G(\mathbf{k})$; (2) diagonalize the Bloch-Hamiltonian matrix, $h(\mathbf{k})$, to obtain all the eigenvalues, $E_n(\mathbf{k})$, and their corresponding eigenvectors, $\psi_{n,p_z}(\mathbf{k})$; a  matrix, $W(\mathbf{k})$, is constructed from these eigenvectors ordered by their eigenvalues $E_n(\mathbf{k})$; (3) check the validity of Eq. (\ref{C6}); (4) construct the matrix $\overline{U}_g(\mathbf{k}) =  W^\dagger(\mathbf{k})U_g(\mathbf{k})W(\mathbf{k})$; and (5) determine the traces of each diagonal block  $\overline{U}_g(\mathbf{k})$ and compare them to the character values of the symmetry group $G(\mathbf{k})$. The electronic energy bands at high-symmetry $\mathbf{k}$-points in Fig. \ref{Fig2} where labeled using this 5-step procedure.

%\bibliographystyle{apsrev4-1}
%\bibliography{bibliography}

\end{document}